\documentclass[12pt,hidelinks]{article}
\usepackage[body={17.5cm, 22cm},right=2cm]{geometry}
\usepackage{color}
\usepackage{amssymb,graphicx}
\usepackage{amsmath}
\usepackage{epsfig}
\usepackage[bookmarks=true,pdfborder={0 0 0}]{hyperref} 
\usepackage{cite}

\allowdisplaybreaks
\begin{document}
\setcounter{page}{0}
\thispagestyle{empty}

\parskip 3pt

\font\mini=cmr10 at 2pt

\def\codename{{\sc SusyHD}}

\begin{titlepage}
~\vspace{1cm}
\begin{center}

{\LARGE \codename: Higgs mass Determination in Supersymmetry  }

\vspace{1cm}

{\large \bf
Javier Pardo Vega$^{a,b}$ and Giovanni Villadoro$^{a}$}
\\
\vspace{.6cm}
{\normalsize { \sl $^{a}$ Abdus Salam International Centre for Theoretical Physics, \\
Strada Costiera 11, 34151, Trieste, Italy}}

\vspace{.3cm}
{\normalsize { \sl $^{b}$ 
SISSA International School for Advanced Studies and INFN Trieste, \\
Via Bonomea 265, 34136, Trieste, Italy }}

\end{center}
\vspace{.8cm}
\begin{abstract}
We present the state-of-the-art of the effective field theory computation of the MSSM Higgs mass,
improving the existing ones by including extra threshold corrections.
We show that, with this approach, the theoretical uncertainty is within 1~GeV 
in most of the relevant parameter space. We confirm the smaller value of the Higgs mass found
in the EFT computations, which implies a slightly heavier SUSY scale. We study
the large $\tan\beta$ region, finding that sbottom thresholds might relax the upper bound on the scale of SUSY.
We present \codename, a fast computer code that computes the Higgs mass and its uncertainty for any SUSY scale,   from the TeV to the Planck scale, even in Split SUSY, both in the $\overline{\rm DR}$ and in the on-shell schemes.
Finally, we apply our results to derive bounds on some well motivated SUSY models, in particular we show how the value of the Higgs mass allows to determine the complete spectrum in minimal gauge mediation.

\end{abstract}

\end{titlepage}

\tableofcontents

\section{Introduction}

With the recent discovery of the Higgs boson the last missing piece of the Standard Model (SM) has been unveiled and all the parameters of the theory measured. The success of the Standard Model in describing all the measured observables at colliders contrasts with the failure of the theory to explain some non-collider observations, such as Dark Matter, matter-antimatter asymmetry, etc.

Among the various completions of the Standard Model proposed so far, supersymmetric (SUSY) theories remain the most attractive option. Not only they screen the electroweak (EW) scale  from ultraviolet (UV) sensitivity to new physics thresholds but they successfully predict the unification of gauge couplings and may provide with a natural WIMP dark matter candidate.

On the other hand, indirect hints for a small hierarchy between the scales of electroweak and supersymmetry restoration (e.g. flavor observables, searches for EDMs, LEP bounds on the Higgs mass), have found stronger support from the recent discovery of a moderately heavy SM-like Higgs \cite{Aad:2012tfa,Chatrchyan:2012ufa} and from the absence of any evidence of superpartners in the first LHC runs at 7 and 8 TeV (see e.g. \cite{susysummary,Craig:2013cxa}).
%TOADD:EXPcitation

While the scale of supersymmetry may still be low (there are still various arguments in favor of this scenario), hopefully within the reach of the next LHC run, it is fair to say that our confidence in predicting the new physics scale based on naturalness arguments weakened substantially \cite{Giudice:2013yca,Arvanitaki:2013yja}. The suspicion that other mechanisms may explain the strength of the weak interactions is becoming stronger and alternative scenarios to low energy SUSY already exist
\cite{ArkaniHamed:2004fb,Giudice:2004tc}. 
It is thus useful to look for different (more-experimentally-driven) methods to infer the scale of the superpartners. One natural candidate is the value of the Higgs mass, which in supersymmetry is calculable in terms of the couplings and the soft SUSY breaking parameters.

In particular, in the Minimal Supersymmetric Standard Model (MSSM) the tree-level Higgs mass is predicted to lie below the $Z$-boson mass up to quantum corrections logarithmically sensitive to the SUSY breaking scale. Therefore the measured value of the Higgs mass gives non-trivial constraints on the spectrum and couplings of the MSSM, allowing to shrink the allowed energy range for the superpartners. 

Given the logarithmic (in)sensitivity of the Higgs mass value to the SUSY scale, high precision is required in such calculation to reliably determine the allowed parameter space of the theory. Besides, the experimental value is now known
with per mille accuracy $m_h=125.09(24)$~GeV \cite{Aad:2015zhl}.  The effort in the Higgs mass calculation has been remarkable, reaching the two-loop and in some cases the three-loop level, with different techniques and schemes, see \emph{e.g.} \cite{Espinosa:1991fc,Hempfling:1993qq,Heinemeyer:1998jw,Heinemeyer:1998np,Espinosa:1999zm,Espinosa:2000df,Degrassi:2001yf,Brignole:2001jy,
Brignole:2002bz,Martin:2002wn,Dedes:2003km,Heinemeyer:2004xw,Kant:2010tf,Giudice:2011cg,Degrassi:2012ry,Draper:2013oza,Bagnaschi:2014rsa}. 
Some of the computations, however, are only valid for small SUSY breaking scales, where log-corrections do not need resummations; in fact currently available computer codes have a very limited range of applicability compared to the allowed parameter space. Moreover, different computations and/or computer codes disagree among themselves, in some cases substantially more than the expected/claimed level of uncertainties.

Given the important role played by the Higgs mass in constraining supersymmetric models, the limitations of the existing codes and the disagreements in the literature, we felt the need to revisit the computation. We put special emphasis on the relevant parameter space to reproduce the experimental value of the Higgs mass, on the study of the uncertainties and on the possible origin of the differences with other methods. In this paper we recompute the Higgs mass in the MSSM using the effective field theory (EFT) approach, which allows to systematically resum large logarithms and to have arbitrary big hierarchies in the spectrum, exploiting the mass gap hinted by the largish value of the Higgs mass and the absence of new physics at the LHC. 

Our computation follows very closely the ones in \cite{Espinosa:1991fc,ArkaniHamed:2004fb,Giudice:2004tc,Giudice:2011cg,Degrassi:2012ry,Draper:2013oza,Bagnaschi:2014rsa}, providing independent checks of such computations.
We improved them in various ways. 
We added all the dominant SUSY threshold corrections including the contributions from bottom and tau sectors, which become important at large $\tan\beta$. In this way we provide the state of the art in the EFT calculation of the Higgs mass. We performed the computation in both the $\overline{\rm DR}$ and the on-shell (OS) schemes, the latter has the advantages of ensuring the correct decoupling limits and keeping the theoretical errors under control in the whole parameter space. We point out that large logarithms arising from splitting the fermions from the scalar superpartners in split SUSY scenarios do not need resummation in the whole region of parameters space relevant for the observed Higgs mass.
We also find that $m_h=125$~GeV may not necessarily bound the SUSY scale to lie below $10^{10}$~GeV (and much below at large $\tan\beta$), but it might extend at arbitrarily high scales. Another outcome of our computation is that, even at maximal stop mixing, the average stop mass is required to be above the TeV scale in order to reproduce the correct Higgs mass in the MSSM. We also performed a study of the various possible uncertainties, showing that for most of the parameter space we are dominated by the experimental ones. We identified some of the sources of disagreement between existing computations/codes. 

We implemented the computation in a new computer code, \codename, which we make public \cite{codeaddr} and which allows to reliably compute the MSSM Higgs mass (and its uncertainties) even when big hierarchies are present in the spectrum. Avoiding slow numerical integrations, the code is fast enough to be used to set the experimental value of the Higgs mass as a constraint on any other SUSY parameter.

Finally we also explore the implications of the Higgs mass on two of the simplest SUSY scenarios:  minimal gauge mediation (MGM) and anomaly mediation. In particular in the first case we show how the value of the Higgs mass allows to determine the complete spectrum of the superpartners.

The paper is organized as follows. In section~\ref{sec:computation} we 
describe our computation of the Higgs mass in the effective field theory approach, 
we study the theoretical uncertainties and we compare our results with the existing ones.
In section~\ref{sec:results} we present the implications of our computation for the SUSY spectrum in different regimes, in particular we show the constraints from the Higgs mass in the parameter range relevant for SUSY searches at hadron colliders, we explored the region of very large $\tan\beta$ and we comment on the (non) importance of extra log resummation when the SUSY spectrum is split. In section~\ref{sec:code} we briefly introduce \codename, a new code to compute the Higgs mass using the EFT technique. 
In section~\ref{sec:applications} we apply our results to two of the simplest SUSY models: minimal gauge mediation and anomaly mediation.  We summarize the most significant results in the conclusions in sec.~\ref{sec:conclusions}. Finally, in appendix~\ref{app:susythr}, we provide the explicit expressions for some of the SUSY thresholds computed in this work and more details about the conventions used in the text. The reader not interested in the technical details of the computation can look directly at secs.~\ref{sec:results} and \ref{sec:applications}.

\section{The computation} \label{sec:computation}

\subsection{The Effective Field Theory technique}
Whenever a theory presents a gap in its energy spectrum 
effective field theory techniques become a very powerful tool.
They exploit the hierarchy of scales to allow a perturbative
expansion in powers of the energy gap. This simplifies the theory
getting rid of irrelevant degrees of freedom and couplings.

Applied to supersymmetry, as the scale of the superpartners
is raised, the Standard Model becomes a better and better EFT,
with corrections from higher dimensional operators decoupling fast, as 
powers of $v/m_{\rm SUSY}$, the ratio between the EW and the SUSY scale.
At leading order in this expansion the presence of supersymmetry at low energy
reduces to a boundary condition for the SM couplings evolved at
the SUSY scale, where they have to match with the full supersymmetric theory.

From the bottom up the technique reduces to taking the measured SM couplings at low energy,
evolving them up to the superpartner scale and matching them to
the full supersymmetric theory living at high scales. The non-trivial relations between the couplings in the supersymmetric theory (in particular between the Higgs quartic, the gauge-Yukawa couplings and the soft terms) translates into a non-trivial condition on the soft SUSY parameters. 
Equivalently one can leave the physical Higgs mass as a free parameter 
to be determined as a function of the UV SUSY parameters. 
Imposing the physical value for the Higgs mass then gives the constraint. 

The use of this technique in the computation of the Higgs mass in the MSSM is quite old \cite{Espinosa:1991fc}, 
however its utility in natural SUSY spectra was limited since corrections 
from higher dimensional operators could not be neglected in that case. 
These techniques became more popular with the advent of Split SUSY scenarios \cite{ArkaniHamed:2004fb,Giudice:2004tc}
and the recent LHC results \cite{Degrassi:2012ry,Arvanitaki:2012ps,ArkaniHamed:2012gw,Draper:2013oza,Bagnaschi:2014rsa}. 

In the rest of the paper, unless specified otherwise, the gauge couplings $g_{1,2,3}$, the Yukawa couplings $y_{t,b,\tau}$ and the Higgs quartic coupling $\lambda$ are assumed to be the SM ones in the $\overline{\rm MS}$ scheme while the soft parameters (masses and trilinear couplings) are in the $\overline{\rm DR}$ or OS schemes. In particular when we refer to our $\overline{\rm DR}$ or OS results it means that the soft masses are $\overline{\rm DR}$ or OS while the couplings are always
taken to be the SM ones in the $\overline{\rm MS}$ scheme.

Our computation is organized as follows:
\begin{itemize}
\item The SM couplings (gauge, Yukawa and quartic) in the ${\overline{\rm MS}}$ scheme are extracted from the corresponding physical quantities at the EW scale at full two-loop level \cite{Buttazzo:2013uya}.  In particular the matching between the top mass\footnote{As usual we interpreted the experimental value $m_t=173.34\pm0.76$\cite{ATLAS:2014wva} as the pole mass, systematic uncertainties coming from this choice can be estimated by rescaling the experimental error on the top mass.} and the top Yukawa coupling is done using full two-loop thresholds plus the leading three-loop QCD one\footnote{Since we do not perform a complete N$^3$LO computation this last correction is also used to evaluate the uncertainties from higher order terms.} from  \cite{Melnikov:2000qh}.

\item The couplings are then evolved from the weak scale to the superpartner scale using the full three-loop renormalization group equations (RGE) for these couplings\footnote{As for the three-loop top Yukawa threshold, the four-loop QCD corrections \cite{vanRitbergen:1997va,Czakon:2004bu} to the strong coupling RGE has been used to estimate the uncertainties.} \cite{Buttazzo:2013uya}.

\item At the SUSY scale the  SM couplings are matched to those of the SUSY theory (converted from either $\overline {\rm DR}$ or OS to the ${\overline{\rm MS}}$ scheme) using the full one loop thresholds (from \cite{Bagnaschi:2014rsa}
and the ${\cal O}(\alpha_{b,\tau})$ corrections from appendix~\ref{app:susythr}) plus the leading two-loop thresholds ${\cal O}(\alpha_s \alpha_t)$ and ${\cal O}(\alpha_t^2)$.
The former is  computed for generic SUSY spectra while the latter (which is generically smaller) is only computed for degenerate scalars. 
\end{itemize}

The final expression for the Higgs mass can thus be written as 
\begin{equation}
m_h^2= v^2 [\lambda(m_t)+\delta \lambda(m_t)]\,,
\end{equation}
where  $v=246.22$~GeV and $\delta \lambda (m_t)$ are the SM threshold corrections (here computed up to two loops) to match the Higgs pole mass to the $\overline{\rm MS}$ running quartic coupling.  The coupling $\lambda(m_t)$ is derived using the RGE and the boundary conditions at the SUSY scale (see below). The RGE for the Higgs quartic coupling are solved together with gauge and Yukawa couplings at three loops, in particular the top Yukawa $y_t$ is extracted from 
\begin{equation}
m_t= \frac{v}{\sqrt2} (y_t(m_t)+\delta y_t(m_t))\,,
\end{equation}
where $\delta y_t(m_t)$ is the SM threshold correction matching the top pole mass with the  $\overline{\rm MS}$ top Yukawa coupling and here computed at NNLO, and N$^3$LO in the strong coupling. The matching at the SUSY scale $Q$ is instead given by
\begin{equation} \label{eq:matchinglambdaHS}
\lambda(Q)=\frac{g^2(Q)+g'{}^2(Q)}{4}\cos^22\beta+\Delta \lambda^{(1)}+\Delta\lambda^{(2)}_{\alpha_t\alpha_s}+\Delta\lambda^{(2)}_{\alpha_t^2}\,,
\end{equation}
where $\Delta \lambda^{(1)}$ contains the 1-loop thresholds matching the Higgs quartic  coupling $\lambda(Q)$ in the $\overline{\rm MS}$-scheme with the one computed in full SUSY in terms of soft terms and couplings, in the $\overline{\rm DR}$ or OS schemes. $\Delta\lambda^{(2)}_i$ are the leading two loop threshold corrections further discussed below.

If some of the superpartners are light compared to the rest of the SUSY spectrum, as in the case of Split SUSY, a new mass threshold develops. In this case two matchings are in order, the first at the Split scale between the SM and the Split SUSY theory, and the second at the SUSY scale, between the Split theory and the MSSM. The evolution up to the Split scale is the same as in the previous case. We then used 1-loop thresholds to do the matchings and 2-loop RGEs to run the Split-SUSY theory \cite{Giudice:2004tc,Giudice:2011cg}. We will show in section~\ref{sec:splitvsHS} that the simplest approach also works in the Split case, i.e. the effect coming from the splitting of the fermions from the scalar superpartners do not need RGE resummation in the parameter region relevant for the observed Higgs mass.

Our computation is very close to the one in \cite{Bagnaschi:2014rsa}, in particular 
we added the contributions from the bottom and tau Yukawas, relevant at large values of $\tan\beta$,
we recomputed the two-loop thresholds ${\cal O}(\alpha_t \alpha_s)$ using the effective potential in \cite{Zhang:1998bm}, and we also included ${\cal O}(\alpha_t^2)$ corrections computed for degenerate scalar masses.

%%%%%%%%%%O(\alpha_t \alpha_s) correction%%%%%%%%%%%%%%%%%%%
The general expression for the two-loop ${\cal O}(\alpha_t\alpha_s)$ corrections is too long to be reported here, but can be accessed through the computer code \codename\ provided in \cite{codeaddr} for $\overline{\rm DR}$ and OS schemes. Our computation in the $\overline{\rm DR}$ scheme agrees\footnote{We thank the authors of \cite{Bagnaschi:2014rsa} for providing the explicit expression of their result for the cross-check.} with the one of \cite{Bagnaschi:2014rsa}. In the limit $m_{Q_3}=m_{U_3}=m_{\tilde t}$ and vanishing gluino mass the OS expression takes the simple form
\begin{align}
\Delta\lambda^{(2)}_{\alpha_t\alpha_s}=
-\frac{y_t^4 g_3^2 }{16\pi^4} \Bigg[\frac52-\frac12\hat{X}_t^2-\left(2-3\hat X_t^2\right) \ln \frac{m_{\tilde t}^2}{Q^2}+3\ln ^2\frac{m_{\tilde t}^2}{Q^2}\Bigg]\,,
\end{align}
while for $M_3=m_{Q_3}=m_{U_3}=m_{\tilde t}$
\begin{align}
\Delta\lambda^{(2)}_{\alpha_t\alpha_s}=
-\frac{y_t^4 g_3^2 }{16\pi^4} \Bigg[4-6\hat{X}_t-4\hat{X}_t^2+\frac34 \hat{X}_t^4
-\left(2-3\hat{X}_t^2\right) \ln \frac{m_{\tilde t}^2}{Q^2}
+3 \ln^2 \frac{m_{\tilde t}^2}{Q^2} 
%-16+ 4\left(2-3\hat{X}_t^2\right) \ln \frac{m_{\tilde t}^2}{Q^2}-12 \ln^2 \frac{m_{\tilde t}^2}{Q^2} 
%+24 \hat{X}_t+16 \hat{X}_t^2 -3\hat{X}_t^4
\Bigg]\,,
\end{align}
where $\hat{X}_t=X_t/m_{\tilde t}$, $X_t=A_t-\mu/\tan\beta$ and $Q$ is the renormalization scale. 
The definition we used for $X_t$ in the OS scheme is given in eq.~(\ref{eq:defXOS}) in appendix~\ref{app:susythr}.

%%%%%%%%%%O(\alpha_t^2) correction%%%%%%%%%%%%%%%%%%%
The two-loop ${\cal O}(\alpha_t^2)$ supersymmetric threshold correction to the quartic coupling can be derived from the corresponding correction to the Higgs mass. We derived it under the simplifying assumption of degenerate scalars while the $\mu$ parameter and the renormalization scale are left free. We used the results in ref.~\cite{Espinosa:2000df} for the ${\cal O}(\alpha_t^2)$ correction to the Higgs mass calculated using the effective potential technique in $\overline{\rm DR}$. Converting in the one-loop ${\cal O}(\alpha_t)$ correction the 
$\overline{\rm DR}$ superpotential top Yukawa coupling and MSSM Higgs vev into the $\overline{\rm MS}$ SM top Yukawa and EW vev  will produce an additional shift contribution at two loops. Analogously for the OS computation there is an extra shift from converting the stop masses and mixings in the one-loop corrections. We subtract the ${\cal O}(\alpha_t^2)$ top-quark contribution because it already appears in the matching at the EW scale. Finally, it is important to notice that there is also a contribution to the matching of the Higgs mass (and the quartic coupling) at the SUSY scale induced by the one-loop contribution of the stops to the wave-function renormalization of the Higgs field, which is instead absent in the ${\cal O}(\alpha_t\alpha_s)$ corrections. 
The complete expression with the details of the calculation can be found in appendix~\ref{app:susythr}, a simplified expression in the OS scheme for the case $\mu=m_{\tilde t}$ and large $\tan\beta$ reads
\begin{equation}
\Delta \lambda_{\alpha_t^2}^{(2)}=\frac{9\, y_t^6}{ (4\pi)^4}
\left[ 3+\frac{26}{3}\hat{X}_t^2 -\frac{11}{6} \hat{X}_t^4+ \frac16\hat{X}_t^6
-\left(\frac{10}{3} - \hat{X}_t^2 \right)\ln\frac{m_{\tilde t}^2}{Q^2}+ \ln^2 \frac{m_{\tilde t}^2}{Q^2} \right]+{\cal O}\left(\tan^{-2} \beta\right).
\end{equation}
%Where $K\simeq0.20$ is defined in appendix~\ref{app:susythr}.
As a cross check we verified analytically that the two-loop ${\cal O}(\alpha_t\alpha_s)$ and ${\cal O}(\alpha_t^2)$ threshold corrections to the quartic coupling (under the assumption of degenerate scalars) cancel the dependence on the renormalization scale of the Higgs mass at the same order.

%%%%%%%%%%O(\alpha_b) correction%%%%%%%%%%%%%%%%%%%
Finally we also included the 1-loop threshold corrections from the bottom (and tau) sector, which are relevant in the large $\tan\beta$ region. The explicit expressions can be found in the appendix~\ref{app:susythr}.
%In particular as discussed in  \cite{Carena:1999py,Carena:2000yi,Draper:2013oza}  we kept the non-linear dependence on
%$\tan\beta$ in the matching of the $\overline{\rm MS}$ bottom Yukawa and 
%$\overline{\rm DR}$ superpotential coupling at the SUSY threshold scale,
%thus resumming the leading $\tan\beta-$enhanced corrections  ${\cal O}( \alpha_b(\alpha_s\tan\beta)^n+\alpha_b(\alpha_t\tan\beta)^n)$. 
At large $\tan\beta$, depending on the size and sign of other parameters, such as the $\mu$ term and the gaugino masses, the net effect is that of reducing the value of the Higgs mass. This effect may even cancel the tree-level contribution and allow for larger SUSY scales (see section~\ref{sec:largetanb}).

The relevance of the supersymmetric thresholds decreases as the SUSY scale increases because of the evolution of the SM running couplings. Among the missing SUSY threshold corrections the most important are the two-loop ${\cal O}(\alpha_t^2)$ when the scalars are not degenerate, the two-loop ${\cal O}(\alpha_t \alpha,\alpha_s\alpha)$ proportional to the electroweak gauge couplings and the three-loop ${\cal O}(\alpha_t\alpha_s^2)$. In the case of large $\tan\beta$ and sizable $\mu$ parameter, the two-loop corrections proportional to the bottom Yukawa can also be relevant, they include the ${\cal O}(\alpha_b\alpha_s, \alpha_b \alpha_t, \alpha_b^2)$ corrections. The contribution of the missing SUSY thresholds to the Higgs mass is estimated to be below 1 GeV even for a spectrum of superparticles as low as 1~TeV, see the next section.

%One important...

\subsection{Estimate of the uncertainties}
\begin{figure} 
\includegraphics[width=1.05\textwidth]{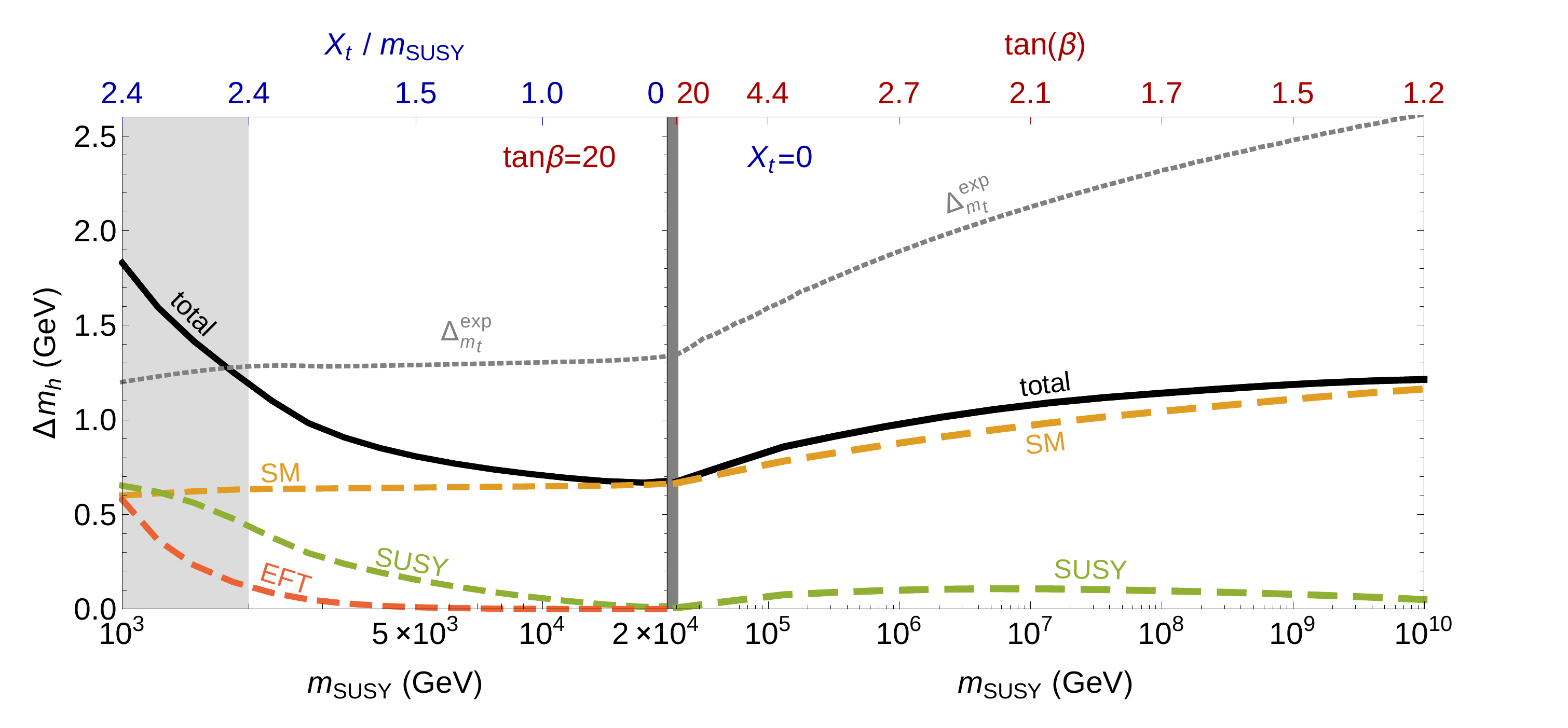}
\caption{\label{fig:uncert} \footnotesize \it Breakdown of the uncertainties for a 125~GeV Higgs mass as a function of the (degenerate) superparticle masses $m_{\rm SUSY}$. 
The Higgs mass has been kept fixed at 125~GeV by varying either the stop mixing (with fixed $\tan\beta=20$ for $m_{\rm SUSY}<20$~TeV, left panel of the plot) or $\tan\beta$ (with vanishing stop mixing for $m_{\rm SUSY}>20$~TeV, right panel of the plot. Note that for $m_{\rm SUSY}<2$~TeV (the gray region) the 125~GeV value for the Higgs mass cannot be reproduced anymore but is within the theoretical uncertainties. The black ``total'' line is the linear sum of the theoretical uncertainties from SM, SUSY and EFT corrections (in dashed lines). The dotted line $\Delta_{m_t}^{\rm exp}$ corresponds to the $2\sigma$ experimental uncertainty on the top mass. }
\end{figure}

In the EFT approach to the calculation of the Higgs mass in SUSY, the uncertainties can be recast into three different groups:
\begin{enumerate}
\item \textit{SM uncertainties}: from the missing higher order corrections in the matching of SM couplings at the EW scale and their RG evolution;
\item \textit{SUSY uncertainties}: from missing higher order corrections in the matching with the SUSY theory at the high scale;
\item \textit{EFT uncertainties}: from missing higher order corrections from higher dimensional operators in the SM EFT and other EW suppressed corrections ${\cal O}(v^2/m_{\rm SUSY}^2)$.
\end{enumerate}
%%%%%%%%%%%%%%%%%%%%%%%%%%%%%%%%%%%%%%%%%
Fig.~\ref{fig:uncert} summarizes the importance of the individual sources of uncertainty as a function of the SUSY scale.
For definiteness we took the superpartners to be degenerate with mass $m_{\rm SUSY}$, the Higgs mass has been kept fixed at 125~GeV by varying either the stop mixing (with fixed $\tan\beta=20$ for $m_{\rm SUSY}<20$~TeV) or $\tan\beta$ (with vanishing stop mixing for $m_{\rm SUSY}>20$~TeV).
We will now discuss these uncertainties individually.

\subsubsection*{SM uncertainties}
As described in the previous section, in our computation we employed full SM three-loop RGE and two-loop matching conditions at the EW scale to relate the pole masses $m_h$ and $m_t$ and the gauge couplings to the $\overline{\rm MS}$ running couplings at the high scale. We also included the 3-loop ${\cal O}(\alpha_s^3)$ corrections to the top mass matching. This is expected to be the leading higher-order correction and the missing 3-loop matching and 4-loop running corrections are not expected to give larger effects.
 Still, we conservatively used the 3-loop ${\cal O}(\alpha_s^3)$ corrections to estimate the SM uncertainties from the higher-order missing corrections, although the latter are probably smaller\footnote{For this reason our theoretical uncertainty from the SM calculation is somewhat larger than the one quoted for example in \cite{Buttazzo:2013uya}, which uses the same precision for the computation of the stability of the SM Higgs potential.}. 

The full SM uncertainty in fig.~\ref{fig:uncert} has been computed by summing the effects from ${\cal O}(\alpha_s^3)$ corrections to the top mass and the $\sim$0.15~GeV estimate \cite{Buttazzo:2013uya} of the 3 loop corrections to $m_h^2$. While the latter corrections would be formally of the same order as the corrections induced by the 2-loop SM corrections to the matching of the top Yukawa in a fixed order computation, in the EFT approach they are actually subleading because the 2-loop corrections from the top sector get RG enhanced. The net effect from these SM corrections amount to a shift to the Higgs mass of order 0.5$\div$1 GeV for $m_{\rm SUSY}\sim 1\div 10^7$~TeV. The uncertainty slowly increases with the SUSY scale as a result of the longer RGE running. 

\subsubsection*{SUSY uncertainties}
\begin{figure} \centering
\includegraphics[width=.4\textwidth]{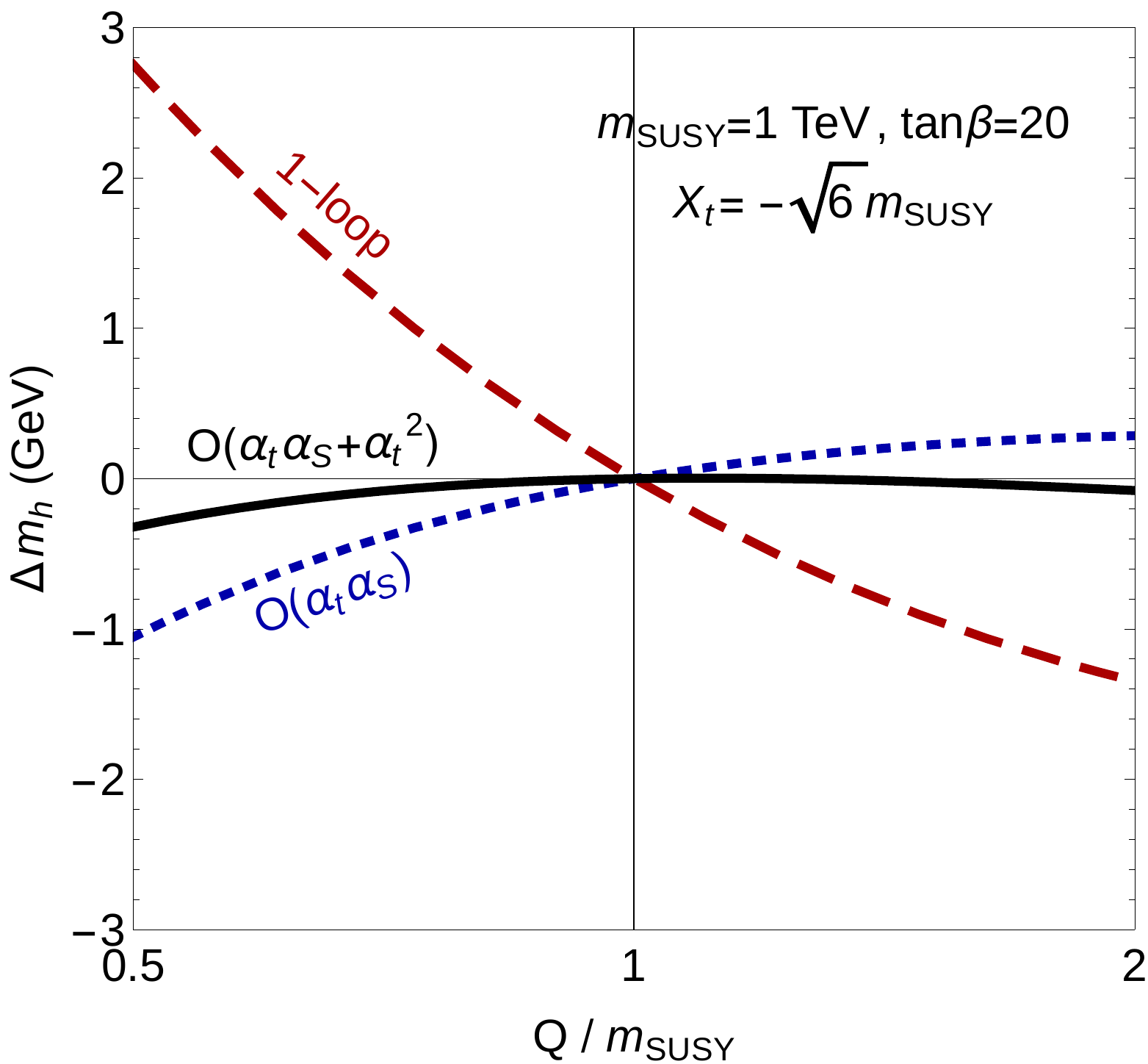}\qquad 
\includegraphics[width=.4\textwidth]{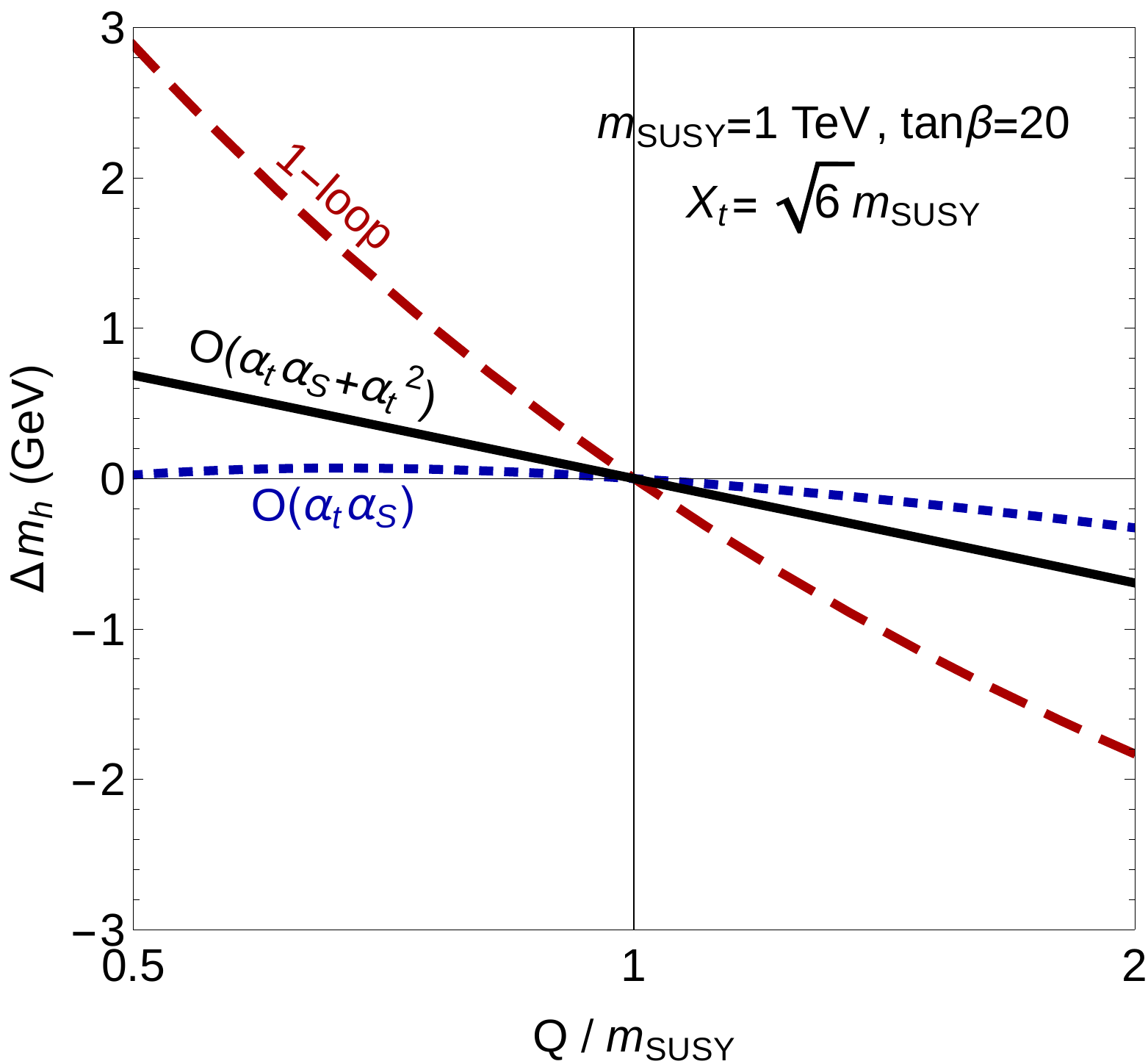}
\caption{\label{fig:scaledep} \footnotesize \it The uncanceled scale dependence from higher order corrections is largest at maximal stop mixing and small stop masses (here taken 1~TeV). Including only 1-loop SUSY threshold it amounts to up to a 3~GeV shift of the Higgs mass, when the scale is changed by a factor of 2. It reduces to  below 1 GeV when the leading 2-loop ${\cal O}(\alpha_t\alpha_s)$ and ${\cal O}(\alpha_t^2)$ corrections are included.}
\end{figure}

The matching between the SUSY soft parameters and the SM couplings includes full one-loop threshold corrections (including also bottom and tau Yukawa corrections) plus the leading two-loop corrections ${\cal O}(\alpha_t \alpha_s)$ and ${\cal O}(\alpha_t^2)$ (the latter only in the simplified case of degenerate scalar masses).
While ${\cal O}(\alpha_t \alpha_s)$ can give large effects, the corrections from ${\cal O}(\alpha_t^2)$  are substantially 
smaller and other missing 2-loop thresholds are expected to be even smaller. Since a missing threshold produce an uncanceled renormalization scale dependence in the final Higgs mass, such dependence can be used to estimate the missing corrections. In the worst case (maximal stop mixing and small SUSY scale) the uncanceled scale dependence from the 1-loop thresholds may shift the Higgs mass by roughly 3~GeV  when the renormalization scale is changed by a factor of 2. ${\cal O}(\alpha_t \alpha_s)$ reduce the shift\footnote{In the right plot of fig.~\ref{fig:scaledep} the scale dependence left after the inclusion of the ${\cal O}(\alpha_t\alpha_s)$ corrections seems to be smaller because of an accidental cancellation for those particular values of the parameters.} to 1~GeV and ${\cal O}(\alpha_t^2)$ further down below 1~GeV, see fig.~\ref{fig:scaledep}.

The uncertainty from SUSY thresholds in fig.~\ref{fig:uncert} has been estimated by taking the maximum of the shifts induced by varying the SUSY matching scale by a factor of 2 or 1/2 with respect to $m_{\rm SUSY}$.
The impact on the uncertainties from missing SUSY threshold corrections greatly reduces away from maximal stop mixing and when the stop masses are increased, the latter effects due to the reduction of the SM couplings from the RGE evolution. This fact is manifest in the EFT approach, less so in others, which require a careful resummation of logs.

\subsubsection*{EFT uncertainties}
The last source of uncertainties is intrinsic to the EFT approach and comes from neglecting higher dimensional operators below the SUSY scale. Such corrections decouple fast, as powers of the EW scale over the SUSY scale, however they could become important for light SUSY scale. Given the relatively high value of the Higgs mass and the bounds from the LHC, superpartners are expected to lie above the  TeV scale, reducing the relevance of these corrections only to very particular corners of the parameter space.
 
 At tree level the only source of power corrections comes from the heavy Higgs states---if $m_A$ is close to the EW scale the mixing effects may become important and the tree-level expression in (\ref{eq:matchinglambdaHS}) gets corrections of order
\begin{equation}
\delta_{EFT} \lambda= -\lambda \frac{m_Z^2}{m_A^2}\sin^2(2\beta)+\dots\,.
\end{equation}
Such contributions become important only when $m_A$  is particularly light ($m_A\lesssim 200$~GeV), in a region of the parameter space which is already disfavored by indirect bounds on the Higgs couplings \cite{atlas:2014-010}.
 
Corrections from higher dimensional operators induced by the other superpartners enter only at one loop (such as the other scalars and the EWinos) or at two loops (gluino). The most dangerous corrections are thus expected to come from the stops, they are of ${\cal O}(\alpha_t\, m_t^2/m_{\tilde t}^2)$ and can get
enhanced at large stop mixing. We estimated such corrections by multiplying the one-loop corrections by $v^2/m^2_{\rm SUSY}$. Numerically, for stops above 1~TeV, even at maximal mixing, these corrections are below 1~GeV and rapidly decouple for heavier stop masses, see fig~\ref{fig:uncert}. Having lighter stops may require to take such corrections into account, although they start being
too light to accommodate the observed value of the Higgs mass (see sec.~\ref{sec:on-shell}). 
Consistently with such approximation we also neglected EW corrections to the sparticle spectrum.

The total EFT uncertainty in fig.~\ref{fig:uncert} has been estimated by taking the sum of the single
contributions to $\Delta\lambda$ from each SUSY particle with mass $m_i$ multiplied by the corresponding
factor $v^2/m_i^2$.

In conclusion, for stops above the TeV scale power corrections are small,  justifying the use of the EFT. 

\subsubsection*{Combined uncertainties}
Fig.~\ref{fig:uncert} summarizes the impact of the various uncertainties to the determination of the Higgs mass as a function of the SUSY scale, $\tan\beta$ and the stop mixing, in the relevant region of parameters that reproduces the measured value of the Higgs mass. For definiteness we took a degenerate spectrum of superpartners, we checked that the size of the uncertainties remains of the same order when this assumption is relaxed.  The dominant source of error comes from higher order corrections in the matching and running of the SM couplings. SUSY thresholds are only important for low SUSY scale and large stop mixing, while power corrections are negligible throughout the parameter space unless some of the sparticles are very close to the EW scale.

It is fair to say that, for most part of the relevant parameter space, the Higgs mass in the MSSM has reached the same level of accuracy as the determination of the Higgs potential in the SM. Further improvements from the theory side can be achieved by extending the SM calculations at higher orders. The size of the uncertainties remains practically unchanged in the split scenario, where the fermions are parametrically lighter than the scalar superpartners.

The total theoretical uncertainty (computed here conservatively as the linear sum of the three sources of errors discussed above) is of order 1~GeV or below for most of the parameter space.
It is thus below the error induced by the experimental uncertainty in the value of the top mass. Indeed, the latter produces a shift in the Higgs mass of order 1.5$\div$2.5~GeV depending on $m_{\rm SUSY}$, when the top mass value is changed by 2$\sigma$=1.5~GeV. The error increases with $m_{\rm SUSY}$ due to RGE effects.

As usual, estimates of theoretical errors provide only for the order of magnitude of the expected corrections and must be taken with a grain of salt. However since for most of the parameter space the error is dominated by the SM uncertainties, where we have been rather conservative, the estimate of fig.~\ref{fig:uncert} should represent a fair assessment, at least away from the lower end.

\subsection{Comparison with existing computations}
Our EFT computation agrees within the uncertainties with all the others which use the same technique. As already noticed in \cite{Bagnaschi:2014rsa}, however, the EFT computation seems to give a smaller Higgs mass with respect to other approaches, such as those based on full diagrammatic and effective potential computations such as \cite{Heinemeyer:1998yj,Allanach:2001kg,Djouadi:2002ze,Porod:2003um}. In some cases the disagreement amounts to up to $\sim$10 GeV, well beyond the expected quoted uncertainties, even in regions of parameter space where both approaches are expected to hold. 

\begin{figure} \centering
\includegraphics[width=.48\textwidth]{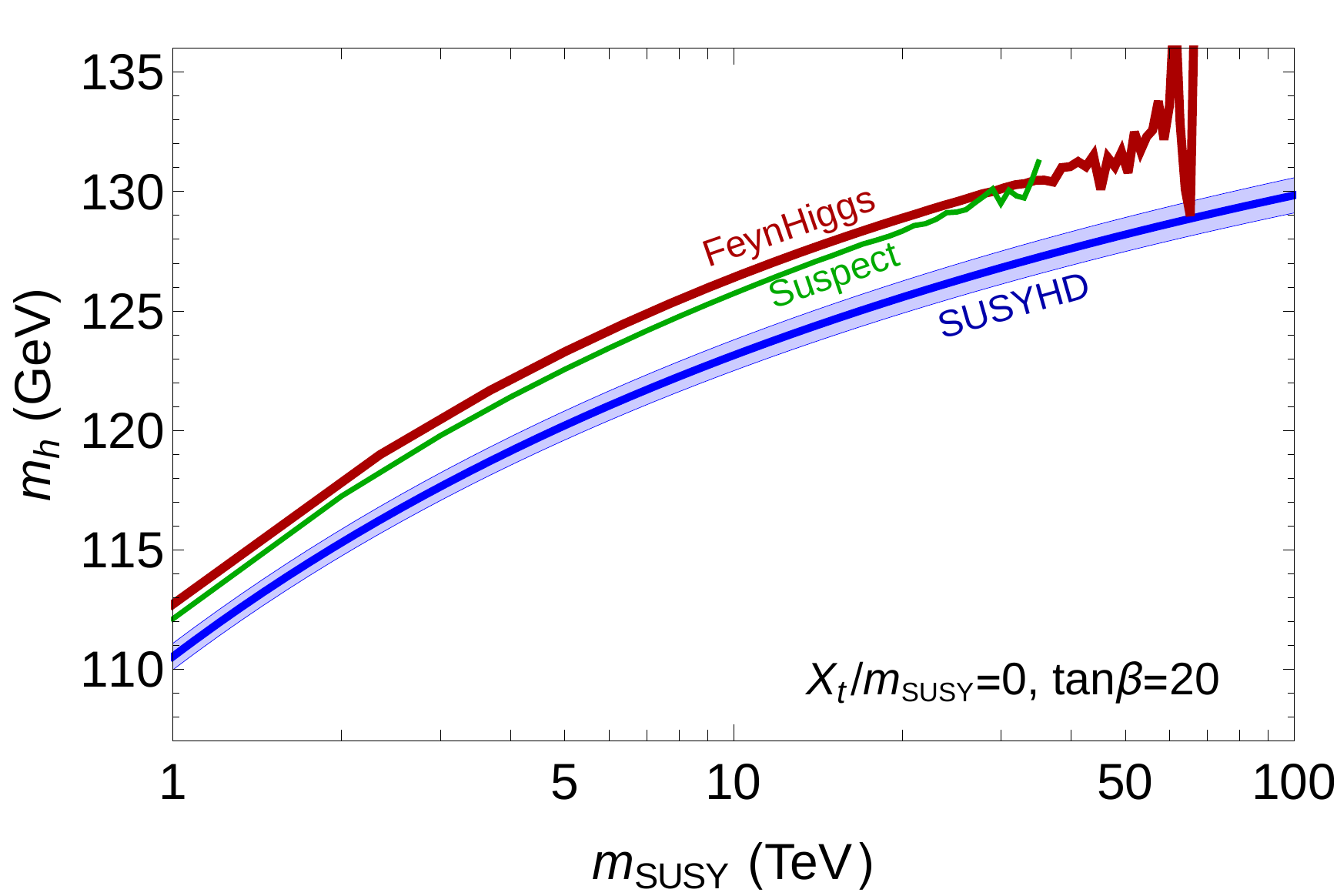}\quad
\includegraphics[width=.48\textwidth]{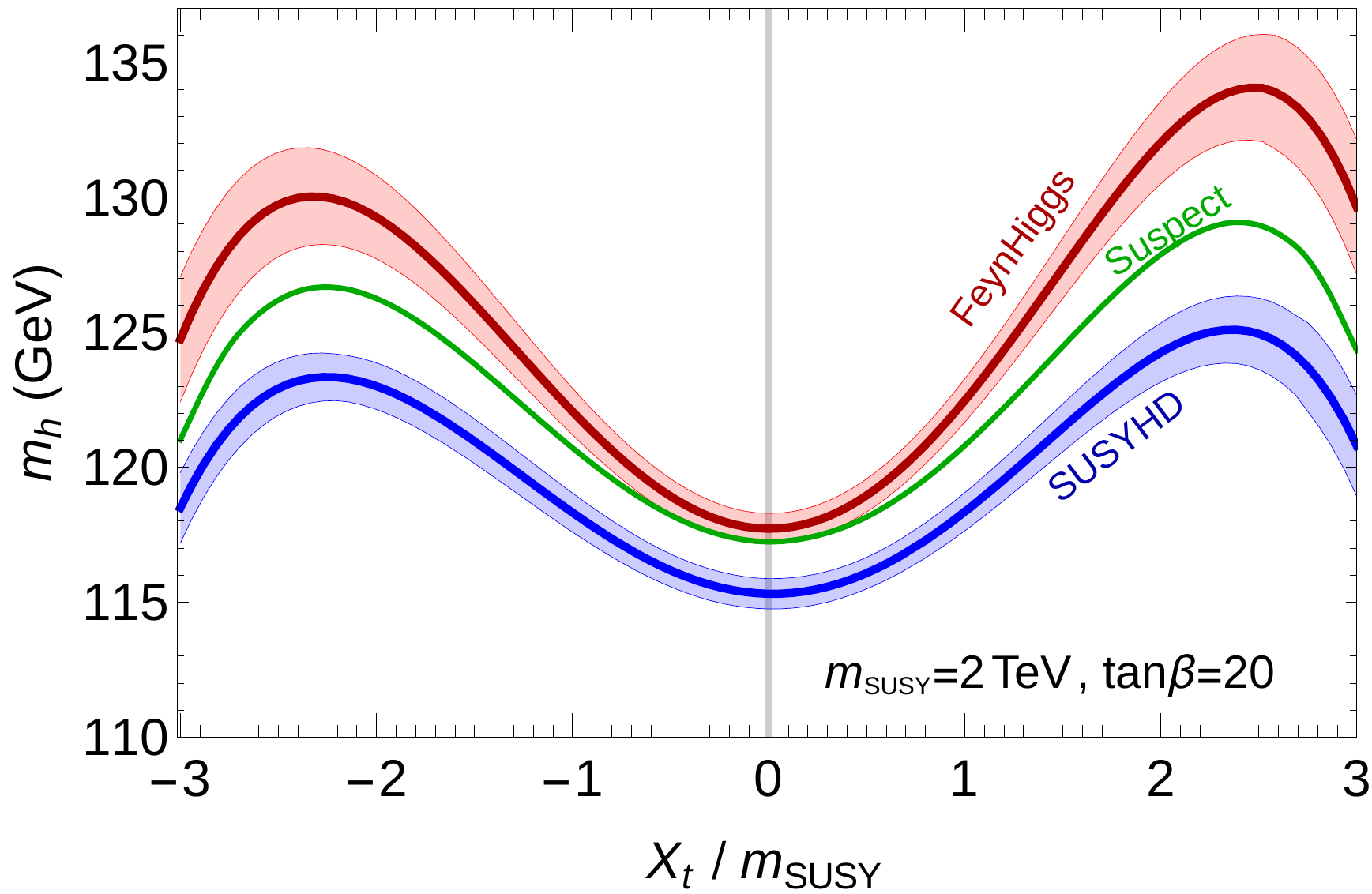}
\caption{\label{fig:comparison} \footnotesize \it Comparison between the EFT computation (lower blue band) and two existing codes: FeynHiggs \cite{Hahn:2013ria} and Suspect \cite{Djouadi:2002ze}. We used a degenerate SUSY spectrum with mass $m_{\rm SUSY}$ in the $\overline{\rm DR}$-scheme with $\tan\beta=20$. The plot on the left is $m_h$~$vs$~$m_{\rm SUSY}$ for vanishing stop mixing. The plot on the right is $m_h$~$vs$~$X_t/m_{\rm SUSY}$ for $m_{\rm SUSY}=2$~TeV.
On the left plot the instability of the non-EFT codes at large $m_{\rm SUSY}$ is visible.}
\end{figure}

A comparison between the EFT computation and some of the available computer codes is shown in fig.~\ref{fig:comparison}.
The disagreement is around 3~GeV for $m_{\rm SUSY}>{\rm TeV}$ at large tan$\beta$ and zero stop mixing
and increases up to 9~GeV for maximal mixing and $m_{\rm SUSY}=2$~TeV.

The large disagreement with the FeynHiggs 2.10.1 code can mostly be understood as follows. 
The computation in \cite{Hahn:2013ria} included full 1-loop plus the leading 2-loop SUSY corrections of the Higgs mass 
with partial 2-loop RGE improvements. 
Consistently with this however, they did not include 2-loop corrections to the matching of the top Yukawa coupling. 
Instead, the use of the N$^3$LO formula shifts the top mass by roughly 4~GeV. 
Hence, the bulk of the disagreement seems due to the missing 2-loop corrections in the top mass\footnote{It was brought to our attention that a similar observation was also made in \cite{slavich:talk}.}. Note that, as discussed in the previous section, the uncertainty in the EFT approach is dominated by the 3-loop top matching conditions, the 2-loop ones are thus mandatory in any precision computation of the Higgs mass.
We checked that after their inclusion, the FeynHiggs code would perfectly agree with the EFT computation at zero squark mixing. At maximal mixing the disagreement would be reduced to 4~GeV, which should be within the expected theoretical uncertainties of the diagrammatic computation.

For comparison, in fig.~\ref{fig:comparison} we also show the results obtained with a different code (Suspect \cite{Djouadi:2002ze}) which uses a diagrammatic approach but unlike FeynHiggs, does not perform RGE improvement and its applicability becomes questionable for $m_{\rm SUSY}$ in the multi TeV region.

\section{Results} \label{sec:results}
After having seen that the EFT computation is reliable for most of the relevant parameter space we present here some of the implications for the supersymmetric spectrum. Given the generic agreement with previous computations using the same approach, we tried to be as complementary as possible in the presentation of our results, putting emphasis on the improvements of our computation and novel analysis in the EFT approach.

\subsection{Where is SUSY?}
\begin{figure}[t!] \centering
\includegraphics[width=.6\textwidth]{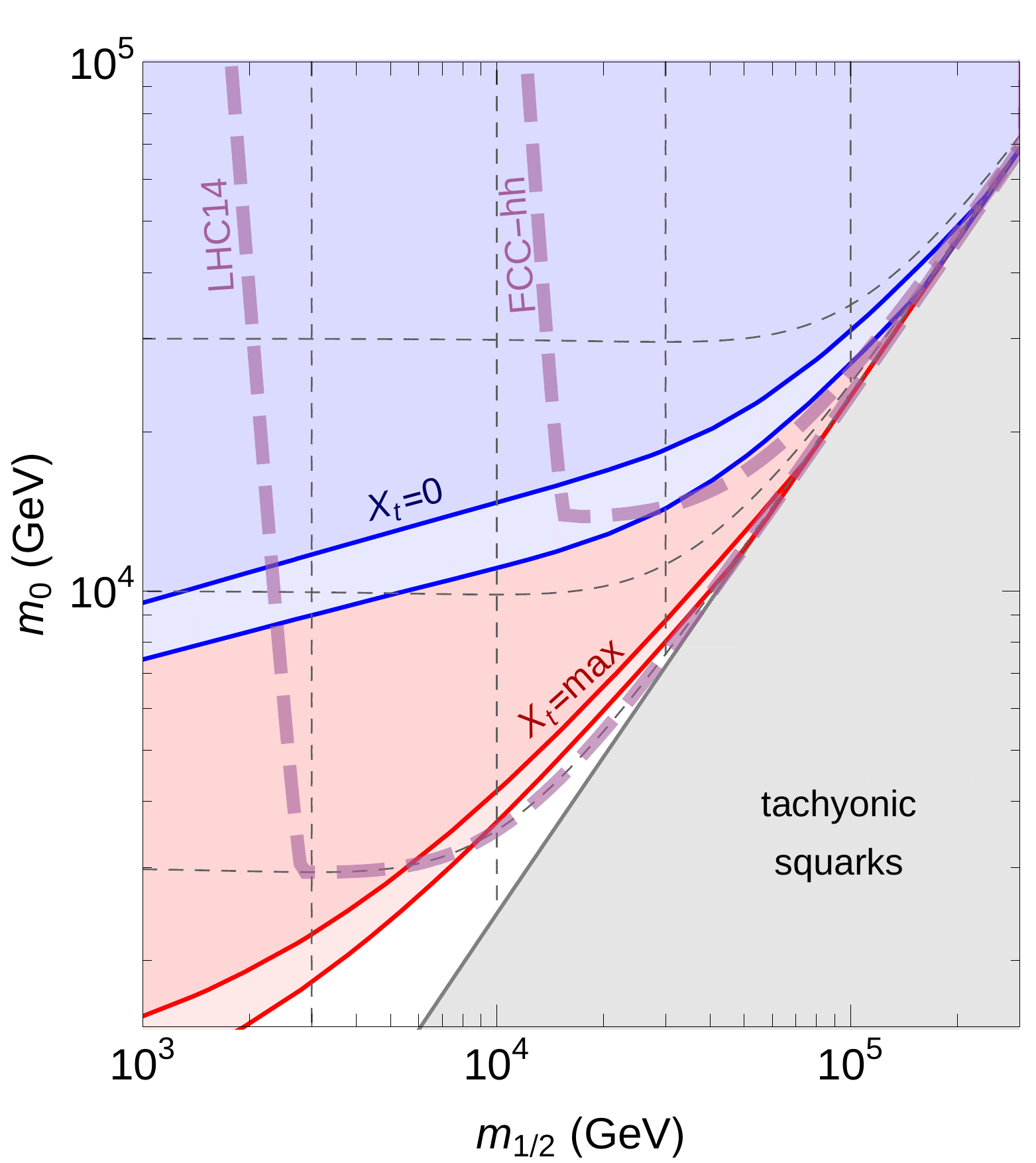}
\caption{\label{fig:m0m12} \footnotesize \it Higgs mass constraint on the value of scalar ($m_0$) and fermionic ($m_{1/2}$) superpartners (taken degenerate). The upper blue region refer to zero mixing, the lower red to increasing values of the stop mixing. The lighter bands corresponds to the uncertainty from the top mass. The gray shaded region corresponds to tachyonic on-shell masses for the squarks.
The non-vertical thin dashed lines correspond to on-shell values for the squark masses: when $m_{1/2}$ grows the $\overline{\rm DR}$ mass $m_0$ must be increased to keep the on-shell mass constant. The thick dashed burgundy lines correspond roughly to the expected reach of $LHC14$ and of an hypothetical 100~TeV machine.}
\end{figure}
Fig.~\ref{fig:m0m12} represents the parameter space compatible with the experimental value of the Higgs mass in the plane of $(m_{1/2},m_0)$ for zero (blue) and increasing values (red) of the stop mixing. For simplicity we took degenerate scalar masses $m_0$ as well as degenerate fermion masses $m_{1/2}=M_{1,2,3}=\mu$. All SUSY parameters of this plot are in the $\overline{\rm DR}$ scheme\footnote{All $\overline{\rm DR}$ parameters are computed at the
scale $Q=\sqrt{m_{\tilde t_L} m_{\tilde t_R}}$ unless specified otherwise.}. The figure highlights a number of features:
\begin{itemize}
\item The main effect at small fermion masses is given by the scale of the scalars (in particular the stops). The lower part of the allowed region corresponds to large values of $\tan\beta\gtrsim 10$. Lowering $\tan\beta$ allows to access larger scalar masses (see also fig.~\ref{fig:largetanb} below). 
\item The dependence on the fermion masses can be understood as follows. 
For $m_{1/2}\lesssim m_0$ the biggest contribution comes from the higgsino-wino loop in the running of the Higgs quartic. It makes the quartic coupling run larger in the IR thus making the Higgs heavier. This correction is only there when both wino and higgsino become light. There is also a smaller 1-loop correction from the individual EWinos, which affects the running of the EW gauge couplings. They make the gauge coupling run bigger in the UV increasing the tree-level contribution to the Higgs quartic (\ref{eq:matchinglambdaHS}) and thus its pole mass. 
Lowering the gluino mass  decreases the Higgs mass
but the effect is two-loop suppressed and only non-negligible at large stop mixings.
The region $m_{1/2}\gtrsim 2 m_0$ should be treated with care. In the $\overline{\rm DR}$ scheme there are negative quadratic corrections to the squark masses proportional to the gaugino masses \cite{Pierce:1996zz}
\begin{equation} \label{eq:squarkOS}
m_{\tilde q}^2{}^{\rm ,OS}=m_{\tilde q}^2{}^{,\overline{\rm DR}}(m_{\tilde q}^2)-\frac{4\alpha_s}{3\pi}M_3^2\left[\log\left(\frac{M_3^2}{m_{\tilde q}^2} \right) -1 \right]+\dots
\end{equation}
In particular when $M_3$ becomes larger than roughly a factor of four with respect to the squark masses the corresponding on-shell masses become tachyonic. Just before this happens the on-shell masses (the dashed lines in the figure) start becoming smaller and smaller with respect to the $\overline{\rm DR}$ parameters, in this tuned region large corrections make the $\overline{\rm DR}$  computation unstable. This explains the strong apparent dependence on $m_{1/2}$ on the right-hand part of the plot, which would disappear if plotted in terms of the on-shell masses. We decided to keep the plot in terms of the $\overline{\rm DR}$ parameters to highlight the tuning required to explore such region.
\item
Current LHC searches already probed squark and gluino masses up to $1.5$~TeV circa \cite{Aad:2014wea}.
This corresponds to the very lowest part of the allowed parameter space, 
where the stop mixing is maximal, $\tan\beta$ is large and fermions must be lighter than scalars. This, of course, with the caveat that the strongest experimental bounds apply to first generation squarks and gluino while the Higgs mass mostly depend on the stops and (somewhat weaklier) on EWino. With the same caveat LHC14 should eventually be able to more confidently explore the same region (extending the squark-gluino reach to $3$~TeV, see e.g.~\cite{Cohen:2013xda}), while the small stop mixing region could only be reached directly with a 100~TeV machine (capable of probing colored sparticles of roughly $15$~TeV masses, see e.g.~\cite{Cohen:2013xda}). Of course (mini-)Split scenarios where the heavy scalars are responsible for the Higgs mass and the light fermions are within reach at lower energies remain a valid possibility.
\end{itemize}

\subsection{The EFT gets on-shell} \label{sec:on-shell}
Previous computations using the EFT approach have used the $\overline{\rm DR}$ scheme for the SUSY and the soft parameters. This scheme has the advantage of being the natural framework for the computations of the soft parameters in theories of SUSY breaking.
In some cases, however, it results inadequate for the computation of the Higgs mass.

First of all, physical on-shell masses are needed to compare theoretical computation with experiments. While the difference in the schemes is one-loop suppressed, there are non-decoupling effects which require care. For example the difference between the on-shell and the $\overline{\rm DR}$ squared mass of the squarks receives an additive one-loop correction proportional to the gluino mass squared, see eq.~(\ref{eq:squarkOS}). Such correction is negative and big---it is enough for the gluino mass to be a factor of four above the squark masses to drive the corresponding on-shell mass tachyonic. 

For similar reasons in the $\overline{\rm DR}$ scheme the gluino contribution to the Higgs mass does not decouple \cite{Degrassi:2001yf}.
 Another consequence is the instability of the Higgs mass with respect to the renormalization scale---even if the on-shell squark masses are positive, the $\overline{\rm DR}$ stop mass becomes highly sensitive to the renormalization scale when the gluino is more than a factor of 2$\div$3 above it, which results in an instability of the estimate of the Higgs mass. What is happening is that the physical on-shell squark masses becomes tuned and highly sensitive to the soft parameters. The situation is similar to trying to compute the Higgs mass in terms of the soft parameter $m_{H_u}^2$ instead of the on-shell (tuned) EW vev $v$.

All these problems disappear in the OS scheme, the gluino decouples up to a physical log correction \cite{Degrassi:2001yf}, there are no tachyons since the physical OS masses are given as input and larger hierarchies can be introduced safely within the SUSY spectrum (with the usual caveat that large logarithms may require resummation). Besides, the input masses are directly the physical quantities to be compared with experiments.
\begin{figure}[t!] \centering
\includegraphics[width=.8\textwidth]{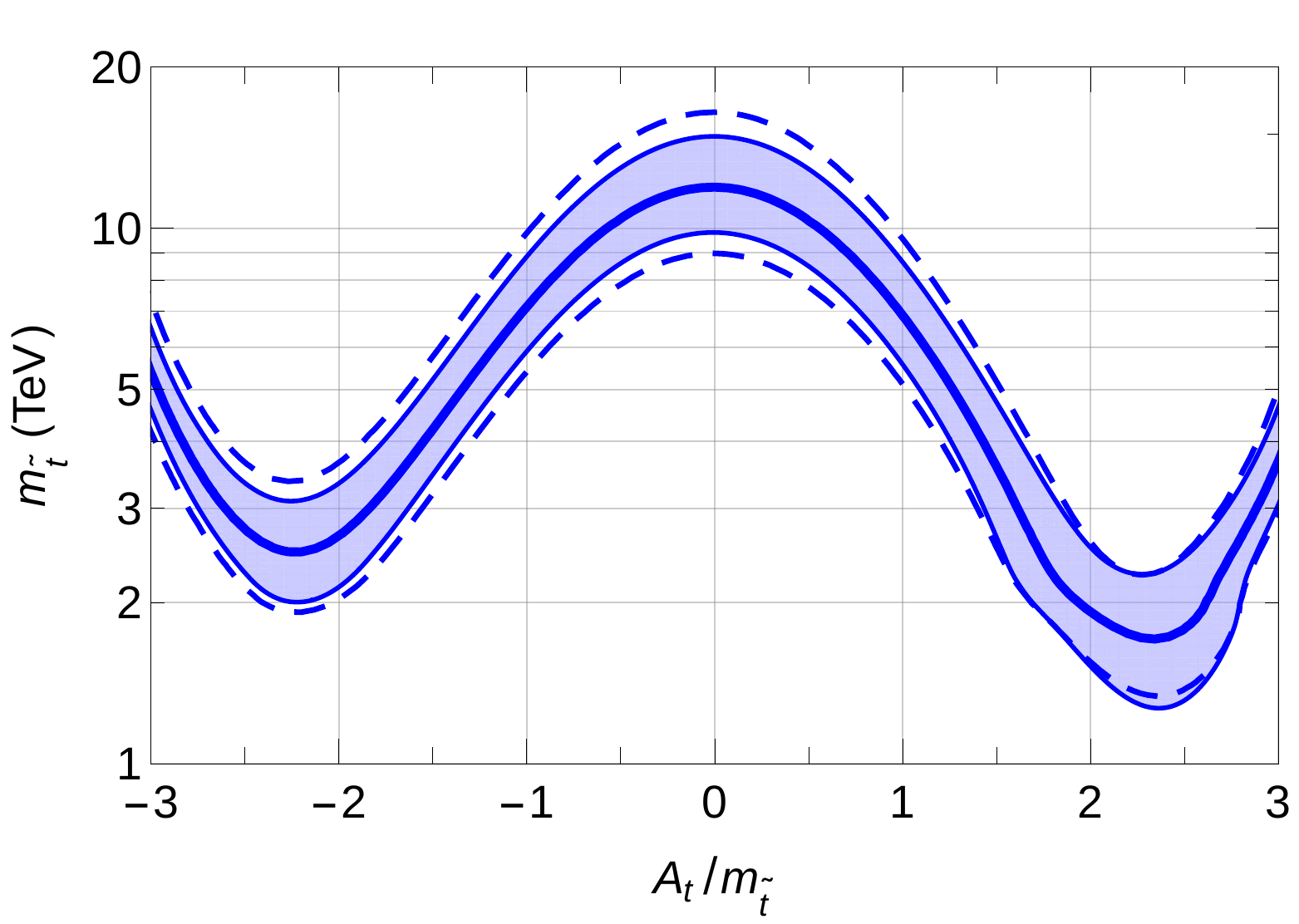}
\caption{\label{fig:mstopAt} \footnotesize \it Allowed values of the OS stop mass reproducing $m_h=125$~GeV as a function of the stop mixing, with $\tan\beta=20$, $\mu=300$~GeV and all the other sparticles at 2~TeV. The band reproduce the theoretical uncertainties while the dashed line the $2\sigma$ experimental uncertainty from the top mass. 
The wiggle around the positive maximal mixing point is due to the physical threshold when $m_{\tilde t}$ crosses $M_3+m_t$. }
\end{figure}

For these reasons we also performed our computation in the OS scheme. Fig.~\ref{fig:mstopAt} shows an application of such calculation. It corresponds to the region of allowed OS stop masses (taken degenerate in this case) which reproduces the observed Higgs mass for different $A_t$-terms. 
Our definition of $A_t$ in the on-shell scheme, eq.~(\ref{eq:defXOS}), is different from the usual one, this explains why the point of maximal mixing is not at $X_t/m_{\tilde t}\simeq2$.
In the spirit of \emph{natural} SUSY \cite{Dimopoulos:1995mi,Cohen:1996vb,Pomarol:1995xc} we kept the higgsino light at 300~GeV while the gauginos and first generation squarks safely above collider bounds at 2~TeV. The lightest stop masses allowed in this case (for maximal stop mixing) are about 1.7$\pm$0.4~TeV, in the region where the EFT approach should be reliable.

Had we drawn the same plot in terms of the $\overline{\rm DR}$ masses we would not be able to draw the same conclusion---the error would blow up in the region where the stops are sufficiently lighter than the gluino.

\subsection{Large-$\tan\beta$ High-Scale SUSY strikes back?} \label{sec:largetanb}
\begin{figure}[t!] \centering
\includegraphics[width=\textwidth]{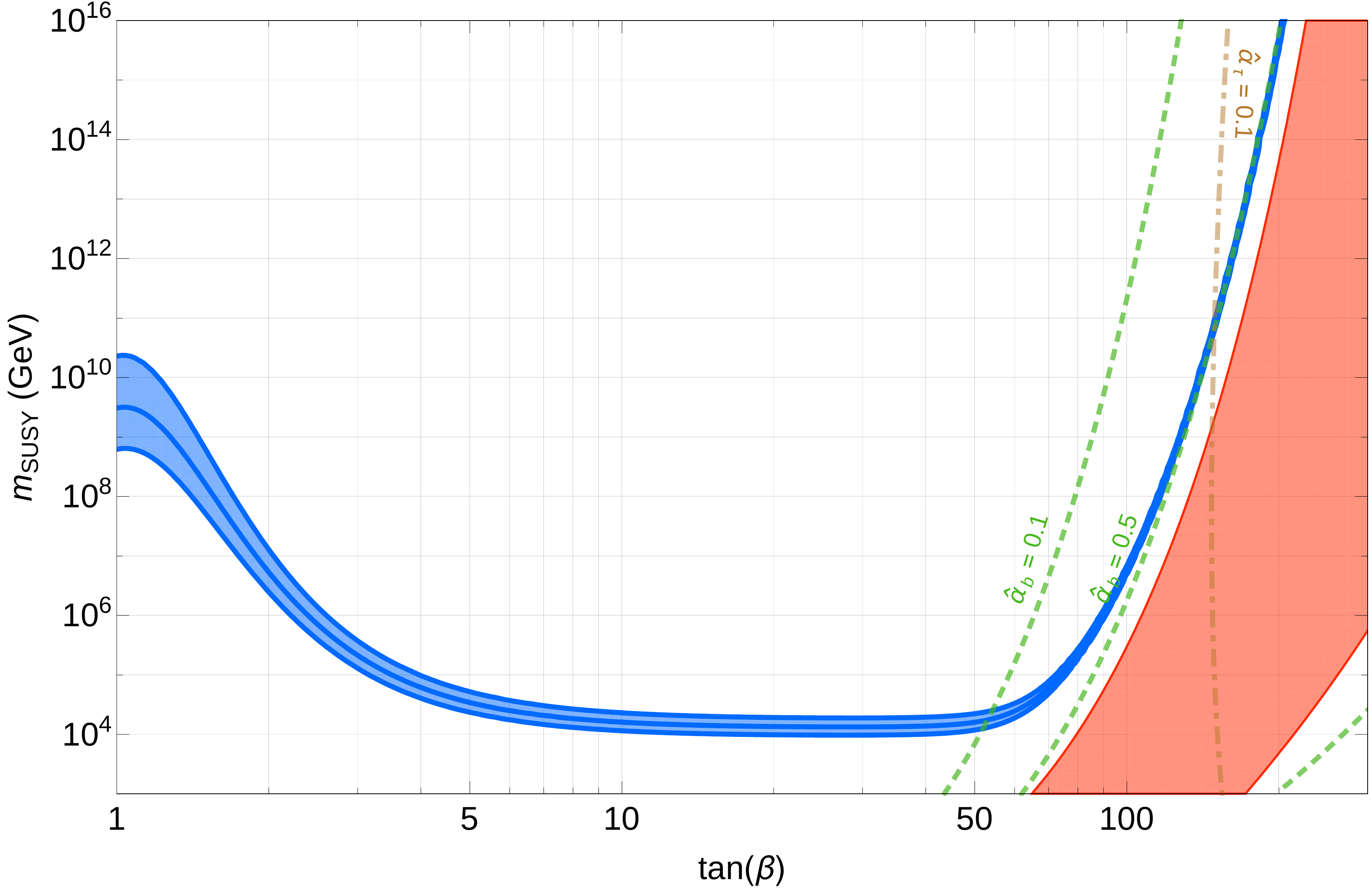}
\caption{\label{fig:largetanb} \footnotesize \it Effects of the sbottom threshold at large $\tan\beta$. 
The blue band corresponds to the $m_h=125$~GeV constraint (the width is given by the estimated theoretical uncertainties) for different values of $\tan\beta$ and the degenerate SUSY mass $m_{\rm SUSY}$. 
We fixed $\mu=-m_{\rm SUSY}$ and $A_t=m_{\rm SUSY}/2$. 
Dashed and dot dashed lines correspond to different values of the bottom and tau couplings respectively.
The red region correspond to tachyonic Higgs and/or non-perturbative bottom Yukawa coupling.}
\end{figure}

While for most values of $\tan\beta$ the contributions from the bottom and tau sector can be neglected, at very large $\tan\beta$ the corresponding superpotential couplings become large and their effects to the SUSY threshold can eventually dominate over the others. In particular the one loop sbottom threshold to the Higgs quartic coupling at leading order in $\tan\beta$ and degenerate sbottoms ($m_{\tilde b}$) reads
\begin{equation} \label{eq:deltab}
\Delta \lambda^{(1)}_{\tilde b}=-\frac{\hat y_b^4}{32\pi^2}\frac{\mu^4}{m_{\tilde b}^4}\,,
\end{equation}
and analogously for the tau. At tree-level the superpotential Yukawa coupling ${\hat y}_b$ is related to the SM Yukawa $y_b$ by ${\hat y}_b=y_b/\cos\beta$. At large $\tan\beta$,  ${\hat y}_b$ may become larger than one. In this situation the negative threshold correction (\ref{eq:deltab}) may cancel or even overcome the tree-level contribution, especially at high SUSY scale where the SM EW gauge couplings are smaller. This effect may allow to maintain $m_h=125$~GeV and large $\tan\beta$ with arbitrary heavy scalar fields, reopening the High Scale SUSY window above $10^{10}$~GeV, which was thought to be excluded by the Higgs mass within the MSSM. As an example, we show in fig.~\ref{fig:largetanb} how the
$m_{SUSY}$-vs-$\tan\beta$ plot would look like at large $\tan\beta$ after including the leading bottom (and tau) contributions.
For the plot we chose degenerate spectrum with mass $m_{\rm SUSY}$, negative $\mu=-m_{\rm SUSY}$ and $A_t=m_{\rm SUSY}/2$. The SUSY parameters are given in the OS scheme. The behavior at small and moderately large $\tan\beta$ ($\tan\beta\lesssim 40$) is well-known \cite{Giudice:2011cg,Arvanitaki:2012ps,Bagnaschi:2014rsa}. However further increasing $\tan\beta$, the bottom coupling $\hat\alpha_{b}=\hat y_b^2/(4\pi)$ grows, 
decreasing the Higgs mass \cite{Degrassi:2002fi}. For very large $\tan\beta$ the tree-level contribution to the bottom mass is so
suppressed that loop corrections cannot be neglected \cite{Banks:1987iu,Hall:1993gn}. In fact the bottom mass receives corrections from SUSY breaking proportional to $v_u=v \sin\beta$, \emph{i.e.} not $\tan\beta$ suppressed
\begin{equation} \label{eq:ybyb}
y_b=\hat y_b \cos\beta+\hat y_b  \sin\beta \left [ \frac83 \frac{\alpha_s}{4\pi} \frac{\mu M_3}{m_{\tilde b}^2}\,F\Bigl (\frac{M^2_3}{m^2_{\tilde b}}\Bigr )
 + \frac{\alpha_t}{4\pi}\frac{1}{\sin^2\beta}\frac{\mu X_t}{m^2_{\tilde b}}\,F\Bigl (\frac{\mu^2 }{m^2_{\tilde b}}\Bigr )\right]+\dots
\end{equation}
where $F(x)=(1-x+x\log x)/(1-x)^{2}$ and we considered $m_{Q_3}=m_{U_3}=m_{D_3}\equiv m_{\tilde b}$. The loop corrections are proportional to $\mu$ and a combination of gauginos and $A$-terms. If the latter are small
or have opposite sign with respect to $\mu$ (like in fig.~\ref{fig:largetanb}), 
$\hat y_b$ will become strong at large $\tan\beta$ in order to reproduce the observed bottom Yukawa (the red region in the plot). Before that, $\hat y_b$ is large enough to make the threshold (\ref{eq:deltab}) win over the tree-level contribution and allow $m_{\rm SUSY}>10^{10}$~GeV at large $\tan\beta$. For example the observed Higgs mass can be reproduced for GUT scale SUSY with $\tan\beta\sim 200$. In this case the bottom coupling $\hat \alpha_b\sim 0.5$, which is still perturbative but with a very close Landau pole $\Lambda_{LP}\approx 10 m_{\rm SUSY}$. The perturbativity of $\hat \alpha_b$ could be improved by choosing larger $\mu$ terms, however this may become in tension with bounds from tunneling into charge/color breaking vacua \cite{Casas:1995pd,Kusenko:1996jn}. 
We do not know what are the corresponding bounds on the $\mu$ term in this regime, this require a dedicated study which is beyond the scope of this work.

%On the right hand side of the red region in fig.~\ref{fig:largetanb}, the negative loop contributions dominate over the tree-level one, which can thus be neglected, the bottom coupling $\hat \alpha_b$ decreases and eventually saturates to the value set by eq.~(\ref{eq:ybyb}). In this region however the $\tau$ coupling is still growing because in the analogous 1-loop expression for $y_\tau$ the leading contributions come from EW corrections only, which are smaller and become important only for much larger $\tan\beta$.

We thus find that the upper bound of $10^{10}$~GeV on the SUSY scale from the observed Higgs mass may not apply for arbitrary values of $\tan\beta$ but only for small to moderately large $\tan\beta$. 
High scale SUSY at larger $\tan\beta$, however, requires large $\mu$ terms, gauginos may be lighter but not too much
since they receive loop corrections. Therefore high scale Split SUSY does not seem possible in this way.

If gaugino masses and/or $A$-terms are large and with the same sign as $\mu$, the loop corrections may saturate the full contribution to the  physical fermion mass. If this happens, arbitrary large values of $\tan\beta$ can be reached without ever running into strong coupling effects.

Finally for smaller $\mu$ (not shown in the plot) the bottom-tau sector remains decoupled from the low energy Higgs, the threshold (\ref{eq:deltab}) is never important and $m_{\rm SUSY}$ at large $\tan\beta$ stays constant. 
The bottom and tau Yukawa couplings still become strongly coupled at large $\tan\beta$ but the effect on the Higgs mass remain small. Of course the effect from the new physics present at the strong coupling scale is model dependent and may be important.

\subsection{Split vs High-Scale SUSY computation} \label{sec:splitvsHS}
\begin{figure}[t!] \centering
\includegraphics[width=.6\textwidth]{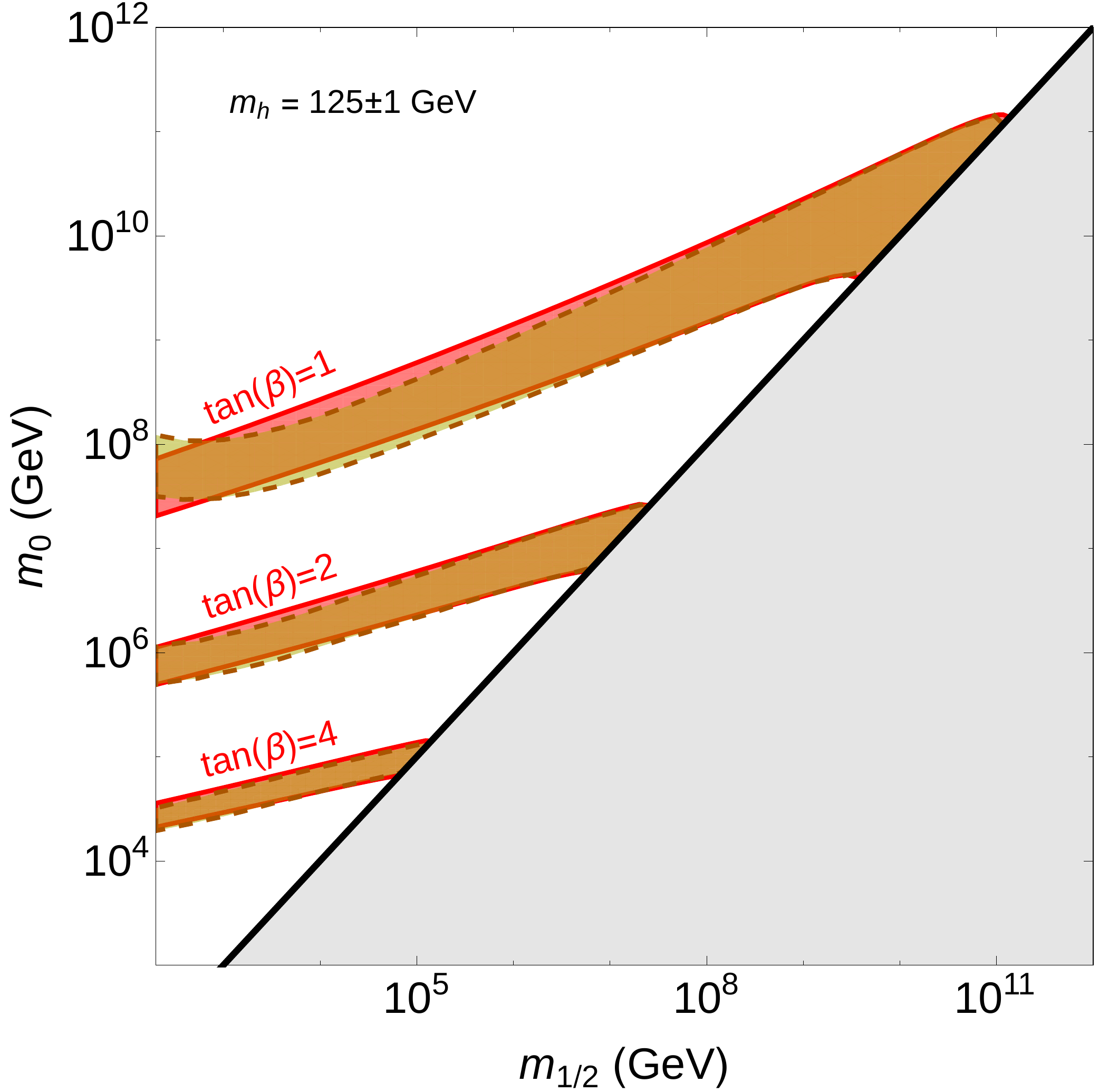}
\caption{\label{fig:HSvsSplit} \footnotesize \it Constraints in the ($m_0$,$m_{1/2}$) plane from $m_h=125\pm1$~GeV for different values on $\tan\beta$ using the proper split-SUSY computation (dark yellow, dashed line) or the approximate high-scale computation (red, continuous line) which does not resum the logs induced by the splitting of $m_{1/2}$ from $m_0$. The agreement is remarkable in the whole relevant parameter space.}
\end{figure}
As mentioned before, in (mini)split-SUSY scenarios, where gauginos and possibly the higgsinos are sensibly lighter than the scalar sector, a new mass scale is present and large logarithms may require resummation. In this case the correct procedure would be: 1) to interrupt the SM running at the split scale, where the light fermion superpartners are, 2) to match to the split-SUSY effective theory, which includes SM particles and the fermion superpartners, 3) to perform a second running within the new EFT and eventually 4) to match to the full SUSY theory at the scalar mass scale. This procedure, which has been employed since the birth of split SUSY, and became more popular recently after the Higgs discovery, is more involved than the high-scale SUSY computation. Besides, the thresholds and the RGE of the split EFT are only known at a lower order in perturbation theory. 

Note however the following. The leading effect of resumming the logarithms generated from splitting the fermion superpartners is to change the running of the Higgs quartic and EW gauge couplings. Numerically the change in the running of the Higgs quartic coupling is the leading contribution but it is only present when both higgsino and gauginos are light. The effect from the change in the RGE of the EW gauge couplings is instead smaller and it is further suppressed at small $\tan\beta$, exactly when the logarithms are the largest.

The observed value of the Higgs mass is not very large and its value limits the SUSY scale to roughly $10^{10}$~GeV. This scale gets further reduced to $10^{7\div 8}$~GeV if the SUSY fermions are split, as an effect of the extra contribution to the running of the Higgs quartic. This translates into an upper bound on how large the logarithmic thresholds from splitting the fermions can grow. 

It turns out that in the whole parameter space relevant for the observed value of the Higgs mass, the effect of resumming the logs of the splitting between fermion and scalar superpartners
% and using the two step procedure with the intermediate running within the split SUSY theory 
is negligible and the results obtained with the single-scale SUSY theory are reliable. 
This is shown in fig.~\ref{fig:HSvsSplit}
where we compare the computation made resumming the logs of the split threshold, using an intermediate Split-SUSY EFT, with the one that does not resum the logs, which uses one scale only and just the SM RGE up to the scalar masses. 
The agreement between the two procedures is impressive. In the worst case ($\tan\beta=1$, fermions at $200$~GeV and scalars at $10^{8}$~GeV) the mismatch is less than 1~GeV, well within the estimated uncertainties. It can also be seen that the two procedures start deviating exactly at that point.
Indeed, had the SUSY scale and the splitting between fermions and scalars been bigger, the two computations would start deviating sensibly, fortunately that region is not relevant for $m_h=125$~GeV.
%independently of the size of the splitting and the scale of SUSY, provided the observed value for the Higgs mass is reproduced.
%
%
%The largest logs are obtained for small $\tan\beta$ when the scalar are the heaviest. In this case however the impact of the threshold is suppressed by RGE effects that decrease all couplings, and the corresponding threshold effects as well. So as the spectrum get more and more split, the logs grows but their coefficient decreases logarithmically as result of the heavier scalar masses. This effect delay the scale where resummation is required. The resulting effect is that log-enhanced corrections from the splitting of SUSY never grow as much---the calculation using only one SUSY scale and the SM EFT provide for a very good estimate for all relevant parameter scale.
%

We conclude that for all the relevant parameter space the computation of the Higgs mass can reliably be made using only the SM as EFT up to the scalar masses, independently of the scale of the fermions\footnote{The usual caveat from $v/m_{\rm SUSY}$ corrections applies when the fermions are very close to the EW scale. In this case the full contributions from the SM+fermion states \cite{Binger:2004nn,Bernal:2007uv} should be used in the matching at the low scale.}, whose main effect is well approximated by the one-loop thresholds at the SUSY scale.

\section{The \codename\ code} \label{sec:code}
The computation described in the previous sections has been implemented into a simple Mathematica \cite{Mathematica} package, \codename\ (SUperSYmmetric Higgs mass Determination), which we made public \cite{codeaddr}. The package provides two main functions that compute the Higgs mass and its theoretical uncertainties from the input soft parameters, and an auxiliary function to change the SM parameters ($m_t$ and $\alpha_s$).

The most time consuming part of the EFT calculation is the integration of the RGE. The code avoids such step by using an interpolating formula for the solution of the RGE, which is only function of the amount of running $\log(Q/m_t)$ and the value of the Higgs quartic coupling at the high scale $\lambda(Q)$, set by the SUSY threshold corrections, eq.~(\ref{eq:matchinglambdaHS}). The interpolating formula only depends on the SM parameters, so the RGE integration needs to be run only once, when the package is first called or if the SM parameters are changed. The result is a very fast code which allows to effectively use the observed value of the Higgs mass as a constraint for the SUSY parameter space. All plots of this paper have been generated with \codename.

The input SUSY parameters can be given in either $\overline{\rm DR}$ or OS schemes and thanks to the EFT approach they can be arbitrary heavy. 
The code also accepts simplified input where not all the SUSY parameters needs to be specified.
There are also extra options which allow: 1) to switch off independently some of the higher order corrections, 2) to change the matching scale $Q$, 3) to use the full numerical code, which integrates the RGE numerically and 4) to use the Split SUSY code which integrates the RGE in two steps: SM up to the fermion scale and Split-SUSY up to the scalars. The function that computes the theoretical uncertainties accepts also the option to compute the individual uncertainties coming from the SM corrections, the SUSY thresholds, and the EFT approximation.

All the necessary documentation can be downloaded with the code from \cite{codeaddr}.

\section{Phenomenological Applications} \label{sec:applications}

\subsection{Predicting the spectrum of Minimal Gauge Mediation}

Gauge mediated supersymmetry breaking (GMSB) \cite{Dine:1981za,Dimopoulos:1981au,AlvarezGaume:1981wy,Nappi:1982hm,Dine:1993yw,Dine:1994vc,Dine:1995ag,Giudice:1998bp} is among the simplest and 
most elegant calculable mechanisms for generating the MSSM soft terms. 
A very special property is the absence of dangerous FCNC, 
a very rare property in extensions of the SM Higgs sector, 
supersymmetric and non.

However, using GMSB to implement a natural solution of the hierarchy problem
has always been hard. The main obstruction being the $\mu$ problem, \emph{viz.} 
why the supersymmetric higgsino mass happens to be at the same scale of the SUSY-breaking soft terms.
Solutions of the $\mu$ problem generically produce a $\mu/B\mu$ problem \cite{Dvali:1996cu}:
both $\mu$ and $B\mu$ are generated radiatively at the same order in perturbation theory,
which produces an unwanted hierarchy, $B\mu \gg \mu^2$. Solutions to the $\mu/B\mu$ problem
exist (see e.g.~\cite{Dvali:1996cu,Giudice:1998bp,Giudice:2007ca} and references therein)
but at the cost of an excessive model building.

All these problems arise when we try to obtain a natural SUSY spectrum---it is like 
we are forcing the theory to do something it was not meant to.
In line with what discussed in the introduction, we are then going to relax 
this requirement and try to use experiments instead of naturalness to infer 
the properties of new physics. 

The apparent gap between the EW and the new physics scale motivates us to revisit
the simplest and more elegant GMSB model, 
minimal gauge mediation (MGM)\footnote{Here by MGM we really mean the most minimal realization,
where the Higgs sector only receives the standard gauge mediated contribution, $\mu$ is a free parameter and $B_\mu$ is generated radiatively in the IR.}, without the unnecessary
baroque model building associated to the Higgs sector. Indeed, ignoring the naturalness problem
allows us to also ignore the $\mu$-problem, as the two are closely related. 
When $\mu$ is much larger than the soft masses EWSB is not possible, 
when $\mu$ is much smaller, the EW scale $v$ would be of order the soft masses.
Therefore if the SUSY scale is above $v$, $\mu$ must automatically be close to the 
SUSY scale in order for EWSB to be tuned to its experimental value.

In MGM all soft masses are generated with the same order of magnitude by the gauge mediated contribution, one gauge loop below the scale $\Lambda=F/M$ (the ratio between the effective scale of SUSY breaking $F$ and the mass of the messengers). Besides $\Lambda$, the spectrum also depends, in a milder way, on the actual mass of the messengers $M$, which determines the amount of running of the soft parameters, and the number of messengers $N$
(typically $N=1$ or 3 for a vector like messenger in the {\bf 5} or {\bf 10} of SU(5) respectively).

As mentioned before, the $\mu$-term, being supersymmetric, would be an independent parameter, but its value is fixed by requiring (tuning) the correct EWSB. Finally the $A$-terms and $B\mu$ are generated radiatively from RGE effects. This fact has very interesting consequences \cite{Dine:1996xk,Rattazzi:1996fb}. First, being $A$ and $B\mu$ terms generated at the quantum level from gaugino masses and $\mu$-term implies that the corresponding CP phases vanish, avoiding potentially dangerous bounds from EDMs. Second, small suppressed $A$-terms imply that the stop mixing will never be large, while small $B\mu$ implies large values of $\tan\beta$. 
These two predictions combined with the measured value of the Higgs mass allows to fix also the 
overall scale $\Lambda$, which must then lie at around the PeV scale
 to produce the ${\cal O}(10)$~TeV SUSY scale required by the Higgs mass. 
 The only remaining free parameters are the messenger mass scale $M$ and their number $N$, which affect the properties of the spectrum in a milder way.
\begin{figure}[t!] \centering
\includegraphics[width=\textwidth]{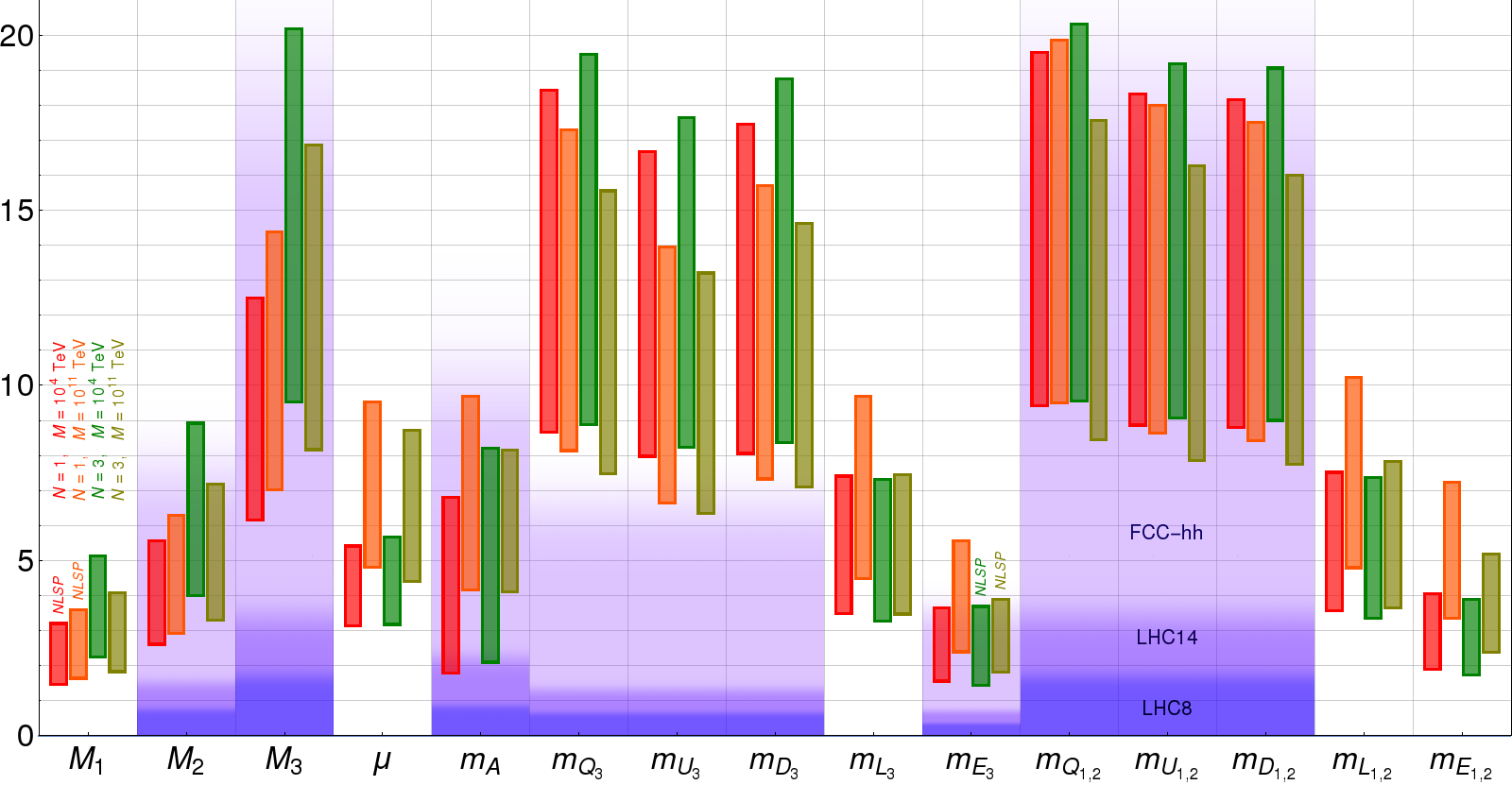}
\caption{\label{fig:MGM} \footnotesize \it 
Prediction for the spectrum of MGM after imposing the constraint from the Higgs mass (or better from the top mass). For each superpartner we plot the allowed range of masses (in TeV) for four different combinations of $N=1(3)$ and $M=10^4(10^{11})$~TeV. For each mass the lowest (highest) value corresponds to increasing (decreasing) the value of the top mass by $2\sigma$ with respect to its experimental central value. The values of $\tan\beta$ at the bottom (top) side of each of the four bands, from left to right, are 58 (42), 49 (45), 56 (29) and 44 (46) respectively.
The three differently shaded areas represent  ``pictorially''  the existing LHC8 bounds and the expected reach at LHC14 and at a future 100~TeV collider, respectively from the bottom.}
\end{figure}

Using our computation for the Higgs mass we can thus \emph{predict} the spectrum of MGM in terms of $N$ and $M$,
the result is shown in fig.~\ref{fig:MGM}. Four different spectra are reported, changing independently
$N$ (1 or 3) and the messenger scale $M$ from $M=10^7$~GeV (to allow the use of leading ${\cal O}(F/M)$ formulae) to $M=10^{14}$~GeV (to avoid dangerous FCNC contributions from gravity mediated contributions).
For each choice of $N$ and $M$ the spectrum is not completely determined because of the uncertainty in the Higgs mass computation. Indeed the effect of varying $N$ and $M$ is actually subleading with respect to the Higgs mass uncertainty. In the relevant region of parameters ($m_{\rm SUSY}\sim 10^4$~TeV and small stop mixing) the Higgs mass determination is at its best (see fig.~\ref{fig:uncert}). Theoretical uncertainties are completely dominated by the SM ones, which are subleading with respect to the experimental uncertainties in the top mass. In fact, what limits the prediction of the MGM spectrum is not the Higgs mass, or its determination in SUSY, but our poor knowledge of the top mass! Improvements in this quantity are required to further improve the predictions of fig.~\ref{fig:MGM}. The lowest (upper) bounds correspond to values of the top mass $2\sigma$ above (below) its measured central value. The overall scale $\Lambda$ results to be at the PeV scale, in particular it varies roughly from 0.5 to 2.6 PeV for different choices of the top mass, $N$ and $M$. The values of $\tan\beta$ are typically around 45 but they can vary up to 60 and down to 30 in the corners of the parameter space, the corresponding values for the supersymmetric bottom and tau Yukawa couplings are largish (typically around 0.5-0.7) but remain always below the one of the top Yukawa. Similarly the stop mixing parameter is always small ${\hat X}_t<1$. 

Except for the overall scale $\Lambda$, which is one order of magnitude larger than the one usually considered in the literature, the rest of the spectrum has the typical GMSB form, with bino or right-handed stau being the NLSP depending if $N=1$ or 3 respectively.

On the experimental side, besides the simplified model and the generic SUSY searches, ATLAS and CMS also performed a number of dedicated GMSB searches \cite{ATLAS:2014-001,CMS:2014koa,Aad:2014gfa,Aad:2014mra,ATLAS:2014fka,Chatrchyan:2013oca,Aad:2014mha,Khachatryan:2014wca}, which exploits some of the most peculiar properties of its spectrum, such as photon and taus in the final states. 
Of particular relevance for this scenario is the direct search 
for the pseudoscalar Higgs boson $A^0$, which, for the large values of $\tan\beta$ predicted here, bounds $m_A\gtrsim 800$~GeV \cite{Khachatryan:2014wca}. 
This channel appears to be the most powerful for MGM, with a slightly better reach than the standard GMSB candles. 

While a dedicated study is required, in fig.~\ref{fig:MGM} we also show ``pictorially'' the existing experimental bounds and the expected reach at LHC14 and at an hypothetical 100~TeV machine. 
The latter are obtained by rescaling the pdf on the existing bounds~\cite{Salam:ColliderReach,diCortona:2014yua} 
and should serve only to guide the eye. However, given the expected scale of the spectrum we can confidently say that this model is mostly out of the reach of existing collider 
machines\footnote{We would like to point out that in some corners of the allowed parameter space $A^0$ may be light enough to be within the reach of LHC14.}, but could be seriously (if not completely) explored by a 100~TeV hadron collider. In fact, MGM may well represent one of the strongest motivations for such machine.

We checked that, while the values of $\tan\beta$ in this model are large, bounds from the rare decays 
$B_{d,s}\to \mu\mu$ \cite{CMS:2014xfa} are not strong enough  to be sensitive to the spectrum in fig.~\ref{fig:MGM} yet.  An improvement on the experimental bounds by a factor 3$\div$5 could be enough 
to start probing the bottom part of the spectrum.

In conclusion, MGM represents probably the simplest and most predictive implementation of SUSY. 
The whole spectrum is almost completely determined just by experimental data.
In particular, the upper bound on the scale of the superpartners exists independently from any naturalness consideration, in fact the value of the Higgs mass predicts a SUSY scale not too far beyond our current reach.  The model also makes other successful \emph{predictions} such as:
gauge coupling unification, the absence of SUSY particles at current hadron colliders, no EDMs 
and no deviation from any flavor observables.
Gravitino may be dark matter although this possibility is more model dependent.

\subsection{Anomaly Mediation}

Minimal anomaly mediated supersymmetry breaking (AMSB) models \cite{Giudice:1998xp,Wells:2003tf}
 probably provide for the simplest implementation of Mini Split supersymmetry. Scalars get their mass from gravity mediation of order the gravitino mass, while fermions, protected by R-symmetry,
get one-loop suppressed soft masses from anomalies. 
The generation of the $\mu$ and $B\mu$ parameters require also the breaking of the PQ symmetry so that the higgsino mass is practically a free parameter---it can be of order the gravitino mass or naturally smaller if the PQ breaking is not efficient.
The theory is thus defined by three main parameters: the gravitino mass $m_{3/2}$ setting the scale of  scalars and gauginos, the higgsino mass $\mu$ and $\tan\beta$ which is determined by the details of the scalar masses and $B_\mu$. 

\begin{figure}[t!] \centering
\includegraphics[width=0.45\textwidth]{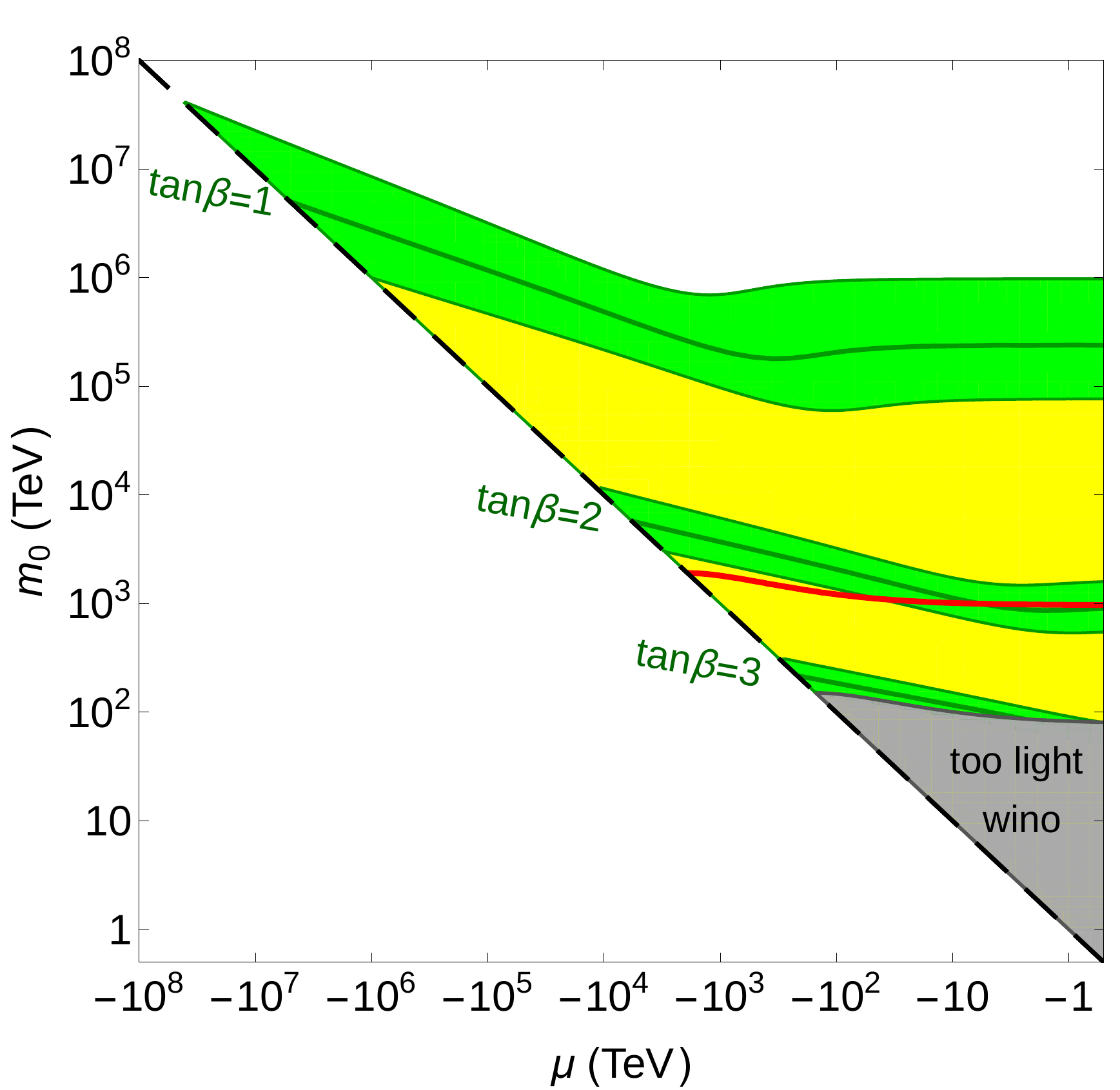}
\includegraphics[width=0.45\textwidth]{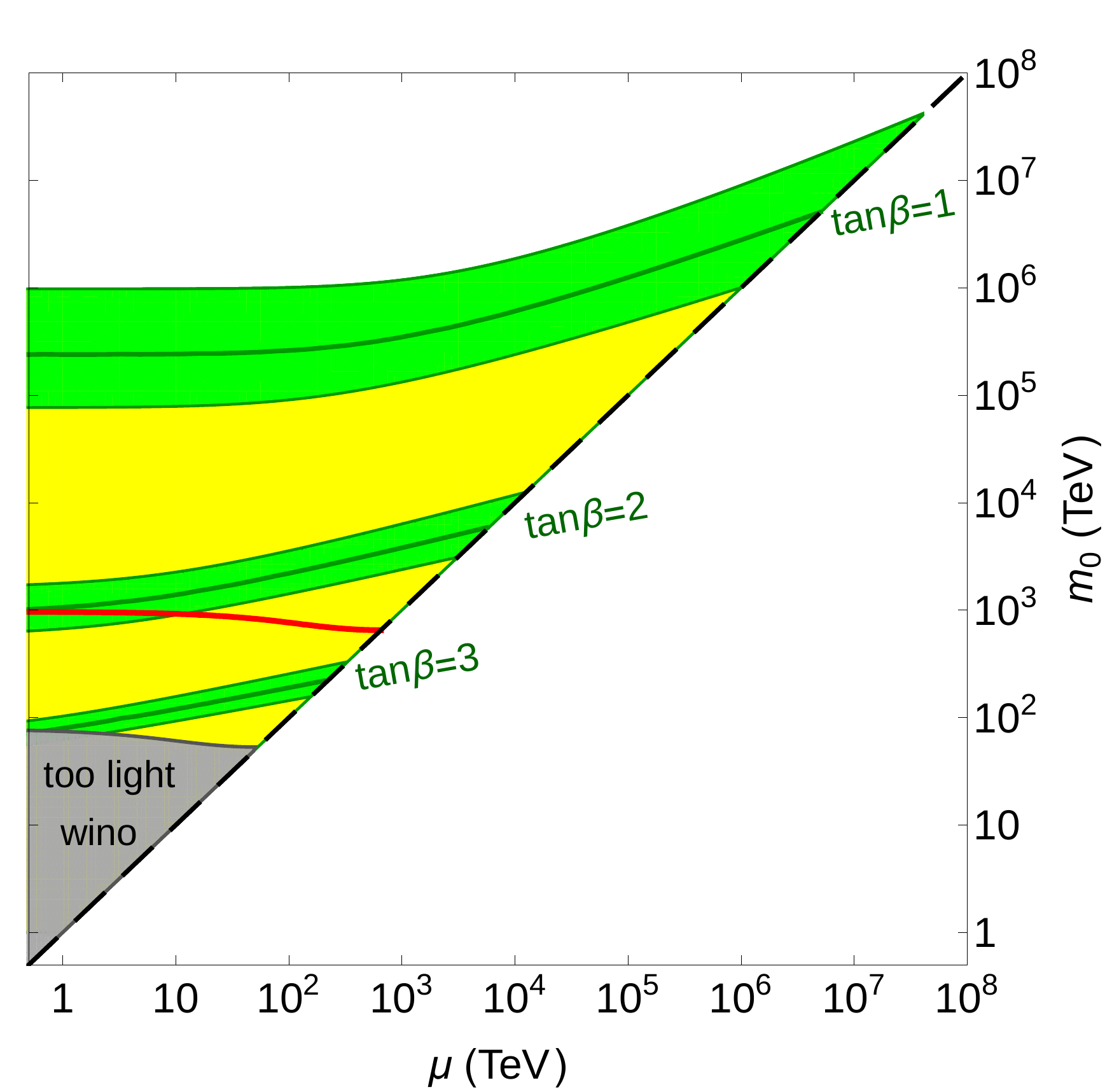}
\caption{\label{fig:AM} \footnotesize \it 
Allowed parameter space of minimal anomaly mediation in the plane ($m_0$, $\mu$) 
for different values of $\tan\beta$, after imposing the constraint from the Higgs mass.
According to AMSB gauginos are one-loop lighter than the scalars, here taken with a common mass $m_0$. The wiggle for negative $\mu$ at small $\tan\beta$ is due to a cancellation in the one loop threshold correction from EWinos when $\mu$ crosses the gaugino masses. Values of $\tan\beta\gtrsim3$ are excluded by LHC bounds on Winos. The horizontal red line corresponds to $M_2\simeq 3$~TeV.}
\end{figure}
Unlike in MGM the details of the scalar spectrum are model dependent, however, given the large scales involved in this scenario, threshold corrections at the SUSY scale are almost irrelevant, for definiteness we fix all the scalars degenerate, $m_0=m_{3/2}$. 
The actual value of the Higgs mass gives a further constraint on these parameters. It can be used, for example, to fix $\tan\beta$ in terms of the other two parameters. It is trivial to impose such constraint using \codename, the result is shown in fig~\ref{fig:AM}. 
Values of $\tan\beta$ larger than 3$\div$4 are already excluded, 
for they would require too low SUSY scale and the wino would lie below the LHC bounds \cite{Aad:2013yna}.
Also a wino with mass $\sim3$~TeV, which would provide for a good thermal dark matter candidate, corresponds to $\tan\beta\sim2\div3$. Bounds on such parameters from direct detection experiment can be found for instance in \cite{diCortona:2014yua}.

Since a LSP wino above $\sim3$~TeV would overclose the universe in the minimal AMSB scenario, the allowed parameter space reduces to the ``narrow'' strip below the red line in fig.~\ref{fig:AM}, and the one with $|\mu|\lesssim1~{\rm TeV}\lesssim M_3$ where the higgsino is the LSP. Most of this parameter space could in principle be probed at a larger hadron collider and future dark matter experiments \cite{Cohen:2013ama,Fan:2013faa,Hryczuk:2014hpa,Low:2014cba,Cirelli:2014dsa,diCortona:2014yua}.
In this interesting region, $\tan\beta$ is constrained between 2 and 3;
scalars are clearly out of reach, between $10^2$ and $10^3$~TeV, but not heavy enough to guarantee the absence of FCNC~\cite{Gabbiani:1996hi}. Gauge coupling unification further prefers values of $\mu$ below ${\cal O}(10)$~TeV \cite{Arvanitaki:2012ps}.

\section{Conclusions} \label{sec:conclusions}

We presented a calculation of the Higgs mass in the MSSM using the EFT approach, which improves previous computations by including extra two-loop SUSY threshold corrections, the contributions from the sbottom/stau sectors relevant at large $\tan\beta$ and the implementation of the OS scheme, the relevant formulae can be found in the appendix~\ref{app:susythr} and in \cite{codeaddr}.

We also performed a study of the theoretical uncertainties, showing that for most of the relevant parameter space the error is sub-GeV and dominated by higher order SM corrections. The result is summarized in fig.~\ref{fig:uncert}. 

The computation has been arranged into an efficient computer package which we made publicly available \cite{codeaddr}. The code exploits the power of the EFT approach, allowing to compute the Higgs mass for arbitrary heavy sparticles, even when a large hierarchy between fermions and scalars is present. 
Analytic formulae for the solution of the RGE make the code very fast, which allows to efficiently use the Higgs mass as a constraint on the spectrum.

We then performed several studies on the implication of the Higgs mass constraint on SUSY:
\begin{itemize}
\item
In agreement with previous EFT computation we find that the SUSY spectrum needs to be a little heavier than expected, 
in particular stops below 2~TeV are disfavored (see fig.~\ref{fig:mstopAt}).
\item
The upper bound on the SUSY spectrum, which is  ${\cal O}(10^{10})$~GeV (${\cal O}(10^4)$~GeV at large $\tan\beta$),
can actually be relaxed without adding new degrees of freedom. At very large $\tan\beta$, if $\mu$ is not suppressed with respect to the scalar masses, sbottom/stau contributions may reduce the Higgs mass, allowing larger values for the SUSY scale (see fig.~\ref{fig:largetanb}).
\item 
In mini-split SUSY, in the region of parameter space relevant for the Higgs mass, the effect of the thresholds from splitting the fermions from the scalars is completely captured by the leading fixed order one loop corrections 
(see fig.~\ref{fig:HSvsSplit}).
This allows to use the SM as an effective field theory all the way up to the scalar mass scale,
avoiding the need of using an intermediate  split SUSY effective theory.
\item
We point out that the value of the Higgs mass may be used to predict  the spectrum of minimal gauge mediation,
the simplest calculable SUSY model, almost completely. The spectrum of SUSY in this case can thus be bounded
just by experimental data alone without the need of arguments based on naturalness. Interestingly enough the spectrum lies just above the expected reach of LHC14 (see fig.~\ref{fig:MGM}), making it an ideal target for a future 100~TeV hadron machine.
\item
Finally we discuss about the analogous implications for anomaly mediation models, constraining the allowed values of $\tan\beta$ and the scale of SUSY (see fig.~\ref{fig:AM}). 
\end{itemize}

For most of the allowed parameter space the Higgs mass computation is dominated by the experimental uncertainty in the top mass. The theoretical uncertainties instead are mostly dominated by the SM higher order corrections.  
Only for maximal stop mixing and at the lightest possible stop masses
uncertainties from SUSY corrections and from higher-order terms in the EFT expansion may 
become important. Improvements in this region can be achieved by including subleading two-loop threshold corrections 
neglected in this work, such as ${\cal O}(\alpha \alpha_{s,t})$ or ${\cal O}(\alpha_t \alpha_{s}^2)$,
and  the leading  ${\cal O}(v^2/m_{\rm SUSY}^2)$ corrections in the EFT expansion.

\section*{Acknowledgements}

It is a pleasure to thank  P.~Slavich for discussions and carefully reading the manuscript, 
his feedback has substantially improved the clarity of the paper.

\appendix
\section{SUSY thresholds}
\label{app:susythr}

This appendix is dedicated to some analytical expressions of the threshold corrections from integrating out supersymmetric particles that were not written in the body for the sake of readability. We start summarizing our conventions. For the numerical part we used the values $m_t=173.34$~GeV~\cite{ATLAS:2014wva}, 
$\alpha_s(m_Z)=0.1185$~\cite{Agashe:2014kda}, $y_b(m_t)=0.0156$ and $y_\tau(m_t)=0.0100$ \cite{Draper:2013oza}.
The MSSM Lagrangian is written with all the parameters in the $\overline{\rm DR}$ scheme (or the ``OS'' scheme described below), and is matched with the SM Lagrangian with all couplings and masses in the $\overline{\rm MS}$ scheme. For $\tan \beta$ we used the definition of \cite{Bagnaschi:2014rsa}. 
As in the rest of the paper, unless specified otherwise, all the formulae are written in terms of the SM couplings ($g_{1,2,3}$, $y_{t,b,\tau}$ and $\lambda$, or $\alpha_i\equiv g_i^2/(4\pi)$ and $\alpha_{t,b,\tau}\equiv y_{t,b,\tau}^2/(4\pi)$) in the $\overline{\rm MS}$ scheme and the soft parameters (masses and trilinear couplings) in the $\overline{\rm DR}$ or OS schemes.
% The renormalization of the parameters in the one-loop SUSY thresholds to $\lambda$ not only affects the expression of the two-loop corrections, but it also changes the central value and the estimated uncertainty of the Higgs mass. Despite the difference between two schemes is higher order, there can be large higher-order corrections due to non-decoupling effects of mass-independet renormalization schemes like the $\overline{\rm DR}$.

The SUSY-breaking masses for the scalars of the $i$-th generation are denoted by $m_{Q_i}$, $m_{U_i}$, $m_{D_i}$, $m_{L_i}$ and $m_{E_i}$, the soft SUSY-breaking Higgs-squarks cubic couplings are written in terms of the superpotential Yukawas $\hat y_{t,b,\tau}$ as $a_t\equiv \hat y_t A_t$, $a_b\equiv \hat y_b A_b$, $a_b\equiv \hat y_b A_b$ for the stops, sbottoms and staus respectively, while the relative 
signs of the $\mu$~parameter, gaugino masses and $A$-terms are the same as in \cite{Martin:1997ns}, so that the scalar 
mass mixings depend on $X_t=A_t-\mu \cot \beta$, $X_b=A_b-\mu\tan\beta$ and $X_\tau=A_\tau-\mu\tan\beta$.

In this work, we extended the one-loop threshold in eq.~(10) of \cite{Bagnaschi:2014rsa} to include also the $\tan\beta-$enhanced contributions from integrating out sbottoms and staus:
\begin{align}
(4\pi)^2\,\Delta \lambda^{1\ell,\,\phi} =& 
3 y_t^2 \left[y_t^2 + \frac{1}{2} \left(g_2^2-\frac{g_1^2}5\right) 
\cos  2 \beta  \right] \ln \frac{m_{Q_3^2}}{\tilde m^2} 
   +3 y_t^2  \left[y_t^2 + \frac{2}{5} g_1^2 \cos 2 \beta \right] 
\ln \frac{m_{U_3}^2}{\tilde m^2} \nonumber \\
% \sum_{\phi={t, b, \tau}} \Bigg\{3 g_\phi^2 \left[g_\phi^2 + \left(g_2^2 T_{\phi_L}^3+\frac35 g_1^2 \left(T_{\phi_L}^3-Q_\phi \right) \right) 
% \cos  2 \beta  \right] \ln \frac{m_{Q_\phi}^2}{\tilde m^2} \nonumber \\
%    &&+3 g_\phi^2  \left[g_\phi^2 + \frac{3}{5} g_1^2 Q_{\phi}\cos 2 \beta \right] 
% \ln \frac{m_{U_\phi}^2}{\tilde m^2}  \Bigg\}   \nonumber \\
&+\frac{ \cos^2 2 \beta}{300}\, \sum_{i=1}^3\, 
\bigg[3 \left(g_1^4+25 g_2^4\right) 
\ln \frac{m_{Q_i}^2}{\tilde m^2}  +24 g_1^4 \ln \frac{m_{U_i}^2}{\tilde m^2} 
+6 g_1^4 \ln \frac{m_{D_i}^2}{\tilde m^2} \nonumber \\
   & ~~~~~~~~~~~~~~~~~~~~~
+ \left(9 g_1^4+25 g_2^4\right) \ln \frac{m_{L_i}^2}{\tilde m^2}   
+18 g_1^4 \ln \frac{m_{E_i}^2}{\tilde m^2}
 \bigg]  \nonumber \\
 &+\frac{1}{4800}  \bigg[261 g_1^4+630 g_1^2 g_2^2  +1325
 g_2^4  - 4 \cos 4 \beta  \left(9 g_1^4+90 g_1^2 g_2^2+175 g_2^4\right) 
\nonumber \\
&~~~~~~~~~~~ -9 \cos 8\beta  \left(3 g_1^2+5 g_2^2\right)^2 \bigg] 
\ln \frac{m_A^2}{\tilde m^2} -\frac3{16} \left(\frac35 g_1^2 + g_2^2\right)^2 
\sin ^2 4 \beta \nonumber \\
&+\sum_{\phi={ t, b, \tau}} \Bigg\{2 N_c^\phi y_{\phi}^4 r_\phi^4 \widetilde X_\phi \left[\widetilde F_1\left(x_\phi\right)
     -\frac{\widetilde X_\phi}{12} \widetilde F_2\left(x_\phi\right)\right] \nonumber \\
&+ \frac{N_c^\phi}{4} y_\phi^2 r_\phi^2 \widetilde X_\phi \cos 2\beta \left[\frac{9}{10} g_1^2 Q_\phi \widetilde F_3 \left(x_\phi\right) + \left(2g_2^2 T_{\phi_L}^3+\frac35 g_1^2 \left(2T_{\phi_L}^3-\frac32 Q_\phi \right)\right) \widetilde F_4 \left( x_\phi \right) \right] 
\nonumber \\
&-\frac{N_C^\phi}{12}  y_\phi^2 r_\phi^2 \widetilde X_\phi  
\left( \frac35 g_1^2 +g_2^2 \right)\cos^2 2\beta\, \widetilde F_5\left(x_\phi \right)\Bigg\}
~.
\label{eq:thresholdscalars}
\end{align}
 In the last three lines of the equation above we sum over the contributions of the stops, sbottoms and staus, where  
 $T_{\phi_L}^3$ is the third component of weak isospin of the left-handed chiral multiplet to which the sfermions belongs, $Q_\phi$ is the electric charge, $\widetilde X_\phi\equiv\{X_t^2/(m_{Q_3}m_{U_3})$, $X_b^2/(m_{Q_3}m_{D_3})$, $X_\tau^2/(m_{L_3}m_{E_3}) \}$, $N_c^\phi\equiv \{3,3,1\}$ is the color factor,  $x_\phi\equiv\{m_{Q_3}/m_{U_3},m_{Q_3}/m_{D_3},m_{L_3}/m_{E_3} \}$, and 
 $r_\phi\equiv \{ 1, \,\hat y_b\cos\beta/y_b,\,\hat y_\tau\cos\beta/y_\tau\}$. The latter coefficients take into account
the $\tan\beta$ enhanced corrections discussed in sec.~\ref{sec:largetanb} which require resummation,
the explicit expressions can be found e.g. in \cite{codeaddr,Draper:2013oza}.
The loop functions $\widetilde F_n$ are defined in appendix A of \cite{Bagnaschi:2014rsa}. Because of the smallness of the bottom Yukawa coupling, the one-loop ${\cal O}(\alpha_b)$ SUSY threshold  corrections are only sizable for large $\tan\beta$ and $|\mu|\stackrel{>}{_\sim}\sqrt{m_{Q_3} m_{D_3} }$.

We obtained the two-loop ${\cal O}(\alpha_t^2)$ SUSY threshold corrections to the quartic coupling of the Higgs from the corresponding correction to the Higgs mass, under the simplifying assumption of degenerate scalars ($m_{Q_3}=m_{U_3}=m_A=m_{\tilde t}$) while the $\mu$ parameter and the renormalization scale are kept independent. The two-loop  ${\cal O}(\alpha_t^2)$ correction to the Higgs mass from the matching between the MSSM and the SM in the EFT approach can be written as the sum of various contributions:
\begin{equation}
m_h^{2\,(\alpha_t^2)}=m_h^{2\,(\alpha_t^2\text{, EP})}+m_h^{2\,(\alpha_t^2\text{, shift})}+m_h^{2\,(\alpha_t^2\text{, WFR})}-m_h^{2\,(\alpha_t^2\text{, top EP})}.
\label{eq:mh2terms}
\end{equation}
The meaning of the various terms in this equation is explained below. The term $m_h^{2\,(\alpha_t^2\text{, EP})}$ is the contribution from the effective potential in the $\overline{\rm DR}$ scheme, which was calculated by Espinosa and Zhang \cite{Espinosa:2000df}:
\begin{align}
m_h^{2\,(\alpha_t^2\text{, EP})}
&= \frac{3y_t^6 v^2}{(4\pi)^4s^2_\beta}\left\{
9{\ln^2\frac{m_{\tilde t}^2}{Q^2}}-6{\ln\frac{m_t^2}{Q^2}}{\ln\frac{m_{\tilde t}^2}{Q^2}}-3\ln^2\frac{m_t^2}{Q^2}+2\left[3f_2({\hat{\mu}})-3f_1({\hat{\mu}})-8\right]{\ln\frac{m_{\tilde t}^2}{m_t^2}}\right.
\nonumber\\
&+6{\hat{\mu}}^2\biggl(1-{\ln\frac{m_{\tilde t}^2}{Q^2}} \biggr)
-2(4+{\hat{\mu}}^2)f_1({\hat{\mu}})+4f_3({\hat{\mu}})-\frac{\pi^2}{3}
\nonumber\\
&+\left[(33+6{\hat{\mu}}^2){\ln\frac{m_{\tilde t}^2}{Q^2}}-10-6{\hat{\mu}}^2-4f_2({\hat{\mu}})+(4-6{\hat{\mu}}^2)f_1({\hat{\mu}})
\right]{\hat{X}}_t^2
\nonumber\\
&+\left[-4(7+{\hat{\mu}}^2){\ln\frac{m_{\tilde t}^2}{Q^2}}+23+4{\hat{\mu}}^2+2f_2({\hat{\mu}})-2
(1-2{\hat{\mu}}^2)f_1({\hat{\mu}})\right]\frac{{\hat{X}}_t^4}{4}
\nonumber\\
&+\frac{1}{2}s_\beta^2{\hat{X}}_t^6\biggl({\ln\frac{m_{\tilde t}^2}{Q^2}}-1\biggr)+
c_\beta^2\biggl[3\ln^2\frac{m_{\tilde t}^2}{m_t^2}  
+7 {\ln\frac{m_{\tilde t}^2}{Q^2}}-4{\ln\frac{m_t^2}{Q^2}}-3+60K+\frac{4\pi^2}{3}
\nonumber\\
&+\biggl(12-24K-18{\ln\frac{m_{\tilde t}^2}{Q^2}}\biggr){\hat{X}}_t^2
-\biggl(3+16K-3{\ln\frac{m_{\tilde t}^2}{Q^2}}\biggr)(4{\hat{X}}_t{\hat{Y}}_t+{\hat{Y}}_t^2)
\nonumber\\
&+\biggl(-6+\frac{11}{2}{\ln\frac{m_{\tilde t}^2}{Q^2}}\biggr){\hat{X}}_t^4
+\biggl(4+16K-2{\ln\frac{m_{\tilde t}^2}{Q^2}}\biggr){\hat{X}}_t^3{\hat{Y}}_t 
\nonumber \\ 
&+\left(\frac{14}{3}+24K-3{\ln\frac{m_{\tilde t}^2}{Q^2}}\right){\hat{X}}_t^2{\hat{Y}}_t^2
-\left(\frac{19}{12}+8K-\frac12{\ln\frac{m_{\tilde t}^2}{Q^2}}\right){\hat{X}}_t^4{\hat{Y}}_t^2\biggr]\Biggr\}.
\label{eq:EPDRbar}
\end{align}
where $X_t=A_t-\mu \cot \beta$, $Y_t=A_t+\mu\cot\beta$, $c_\beta\equiv \cos\beta$, $s_\beta\equiv \sin\beta$,
we use the notation $\hat z\equiv z/m_{\tilde t}$ where $z$ stands for any of the parameters $\mu$, $X_t$ or $Y_t$, and the definitions
\begin{align}
f_1(\hat \mu)&=\frac{\hat \mu^2}{1-\hat \mu^2}\ln\hat \mu^2,
\\
f_2(\hat \mu)&=\frac{1}{1-\hat \mu^2}\left[1+\frac{\hat \mu^2}{1-\hat \mu^2}\ln\hat \mu^2
\right],
\\
f_3(\hat \mu)&=\frac{(-1+2\hat \mu^2+2\hat \mu^4)}{(1-\hat \mu^2)^2}
\left[\ln\hat \mu^2\ln(1-\hat \mu^2)
+Li_2(\hat \mu^2)-\frac{\pi^2}{6}-\hat \mu^2\ln\hat \mu^2
\right],
\\
K &=-\frac{1}{\sqrt3}\int_0^{\pi/6}dx\ \ln(2\cos x)\simeq -0.1953256.
\end{align}
Here $Li_2(x)$ is the dilogarithm function. Below the SUSY scale we use the SM as an effective field theory in the $\overline{\rm MS}$ scheme. Then  we need to write the MSSM top mass and the EW vev (in the $\overline{\rm DR}$ scheme) in terms of the SM ones in the $\overline{\rm MS}$ scheme in the one-loop ${\cal O}(\alpha_t)$ correction to the Higgs mass. Doing so will produce an additional (shift) contribution at two loops
\begin{align}
m_h^{2\,(\alpha_t^2\text{, shift})}&=\frac{3y_t^6 v^2}{(4\pi)^4s^2_\beta} \bigg\{
\left(-\frac32+3\ln \frac{m_{\tilde t}^2}{Q^2} -6\ln \frac{m_{\tilde t}^2}{m_t^2}\ln \frac{m_{\tilde t}^2}{Q^2}+3\ln \frac{m_{\tilde t}^2}{m_t^2} \right)(1+c^2_\beta)  +3 \hat \mu^2 f_2(\hat \mu)  
\nonumber
\\
& -6\hat \mu^2 f_2(\hat \mu)\ln \frac{m_{\tilde t}^2}{m_t^2}-\frac{\hat X_t^6}{12}s^2_\beta
+\hat X_t^2\left[\left(3-6\ln \frac{m_{\tilde t}^2}{Q^2}\right)(1+c^2_\beta)+s^2_\beta\ln \frac{m_{\tilde t}^2}{m_t^2}  -6\hat \mu^2 f_2(\hat \mu)\right]
\nonumber
\\
& +\hat X_t^4 \left[  \frac34-\frac54 c^2_\beta+\frac12\hat\mu^2 f_2(\hat \mu)+\frac12(1+c^2_\beta) \ln \frac{m_{\tilde t}^2}{Q^2} \right]
\bigg\}
.
\end{align}
Unlike the two-loop ${\cal O}(\alpha_t\alpha_s)$ correction to the Higgs mass, the ${\cal O}(\alpha_t^2)$ one receives  a wave-function renormalization contribution. It arises as a combination of the one-loop ${\cal O}(\alpha_t)$ contribution of the stops to the wave-function renormalization of the Higgs field and the one-loop correction to the Higgs mass from the matching at the SUSY scale. It reads:
\begin{equation}
m_h^{2\,(\alpha_t^2\text{, WFR})}=-\frac{3 \,y_t^6 v^2}{(4\pi)^4}\hat X_t^2  \left(\ln\frac{m_{\tilde t}^2}{m_t^2}+{\hat{X}}_t^2-\frac{1}{12} {\hat{X}}_t^4\right).
\end{equation}
Finally, we need to subtract the ${\cal O}(\alpha_t^2)$ corrections to the Higgs mass associated with the contribution of the top-quark loops to the effective potential because 
it is already present in the matching at the EW scale. The two-loop ${\cal O}(\alpha_t^2)$ correction to the Higgs mass in the SM from the matching at the top mass, which receives EP and WFR contributions, is given in eq.~(20) of \cite{Degrassi:2012ry}. We extract the EP piece which is given by
\begin{equation}
m_h^{2\,(\alpha_t^2\text{, top EP})}=-\frac{3 \,y_t^6 v^2}{(4\pi)^4}\left(2+\frac{\pi^2}{3}-7\ln\frac{m_t^2}{Q^2}+3\ln^2\frac{m_t^2}{Q^2}  \right).
\end{equation}

Evaluating eq.~(\ref{eq:mh2terms}) we obtain for the Higgs quartic coupling
\begin{align}
\Delta \lambda^{(2)}_{\alpha_t^2}=&\frac{3 \,y_t^6 }{(4\pi)^4 s_\beta^2}\Bigg\{
\left(-4\ln \frac{m_{\tilde t}^2}{Q^2} +3\ln^2 \frac{m_{\tilde t}^2}{Q^2}\right)s_\beta^2-6{\hat{\mu}}^2\ln \frac{m_{\tilde t}^2}{Q^2}+
\frac12+6{\hat{\mu}}^2-(8+2{\hat{\mu}}^2)f_1({\hat{\mu}})
\nonumber 
\\ 
&+3{\hat{\mu}}^2 f_2({\hat{\mu}})+4f_3({\hat{\mu}})+{\hat{X}}_t^6 s^2_\beta \bigg(-\frac12+\frac12\ln \frac{m_{\tilde t}^2}{Q^2} \bigg) \nonumber\\
&+{\hat{X}}_t^2\bigg(-7-6{\hat{\mu}}^2+4f_1({\hat{\mu}})-6{\hat{\mu}}^2f_1({\hat{\mu}})-4f_2({\hat{\mu}})-6{\hat{\mu}}^2 f_2({\hat{\mu}})+27\ln\frac{m_{\tilde t}^2}{Q^2}+6{\hat{\mu}}^2\ln \frac{m_{\tilde t}^2}{Q^2}  \bigg)
\nonumber
\\
&+\frac{{\hat{X}}_t^4}{2} \bigg(11+2{\hat{\mu}}^2-f_1({\hat{\mu}})+2{\hat{\mu}}^2 f_1({\hat{\mu}})+f_2({\hat{\mu}})+{\hat{\mu}}^2 f_2({\hat{\mu}})-13\ln \frac{m_{\tilde t}^2}{Q^2}-2{\hat{\mu}}^2 \ln \frac{m_{\tilde t}^2}{Q^2} \bigg)
\nonumber
\\  
&+c^2_\beta\bigg[-\frac{13}{2}+60K+\pi^2+9\ln \frac{m_{\tilde t}^2}{Q^2}+{\hat{X}}_t^2\bigg(15-24K-24 \ln \frac{m_{\tilde t}^2}{Q^2} \bigg)-{\hat{X}}_t^4 \bigg(\frac{25}{4}-6\ln \frac{m_{\tilde t}^2}{Q^2}\bigg)
\nonumber
\\
&-{\hat{X}}_t{\hat{Y}}_t \bigg( 12+64K -12\ln \frac{m_{\tilde t}^2}{Q^2}\bigg)+{\hat{X}}_t^3 {\hat{Y}}_t \bigg(4+16K-2 \ln \frac{m_{\tilde t}^2}{Q^2} \bigg)-{\hat{Y}}_t^2 \bigg(3+16K-3\ln \frac{m_{\tilde t}^2}{Q^2}\bigg)
\nonumber
\\ 
& +\hat X_t^2 \hat Y_t^2 \left(\frac{14}{3}+24K-3 \ln \frac{m_{\tilde t}^2}{Q^2} \right)
+{\hat{X}}_t^4{\hat{Y}}_t^2\bigg(-\frac{19}{12} -8K+\frac12 \ln \frac{m_{\tilde t}^2}{Q^2}  \bigg)
\bigg]\Bigg\}.\nonumber\\
\label{eq:alphat2correction}
\end{align}
After taking into account all the contributions in eq.~(\ref{eq:mh2terms}), we checked that the logarithmic dependence on the top mass of the SM quartic coupling is canceled, as it is shown in eq.~(\ref{eq:alphat2correction}). We also verified analytically that the inclusion of the two-loop ${\cal O}(\alpha_t^2)$ correction in eq.~(\ref{eq:alphat2correction}) makes the result of the pole Higgs mass independent of the renormalization scale at this order. 

The two-loop ${\cal O}(\alpha_t \alpha_s)$ correction was also re-computed in this work. The explicit expressions for the SUSY thresholds are too long to be reported here and can be found in the \codename\ package \cite{codeaddr}.

%%%%%%%%%%%%%%%%%%%%%%%%%%%%%%
\subsection*{On-shell scheme}

A change in the renormalization of the parameters entering in the one-loop SUSY thresholds to $\lambda$ will produce a two-loop (shift) contribution. We present the relation between the MSSM parameters in the $\overline{\rm DR}$ and OS schemes. In particular, we need the relations for the stop masses and mixing at ${\cal O}(\alpha_s)$ and  ${\cal O}(\alpha_t)$, the latter for degenerate stops. This will determine the shift contributions to the two-loop ${\cal O}(\alpha_t \alpha_s)$ and ${\cal O}(\alpha_t^2)$ SUSY corrections in the OS scheme.

In the OS renormalization scheme the masses are defined as the poles of the propagators. The relation between the $\overline{\rm DR}$ and OS masses for a scalar particle with squared mass $m^2$ is given by
\begin{align}
m^{2 \, {\rm(OS)}}&=m^{2 \, (\overline{\rm DR})}(Q)-\delta m^2(Q)
\\
\delta m^2(Q)  &\equiv {\rm Re}\, \hat\Pi(m^2,Q),
\end{align}
where $m^{2 \, (\overline{\rm DR})}(Q)$ is the tree-level $\overline{\rm DR}$ mass evaluated at the renormalization scale $Q$, and  $\hat\Pi(m^2,Q)$ is the 
$\overline{\rm DR}$ renormalized one-loop self-energy.

On the other hand, the OS renormalization for the mixing angle is more subtle. At tree-level, the mixing angle of the stops is
\begin{equation} \label{eq:defmix}
 \sin 2 \theta_{\tilde t}=\frac{2m_t X_t}{m_{\tilde t_1}^2-m_{\tilde t_2}^2}.
\end{equation}
We use the symmmetric renormalization for the stop mixing angle (for a discussion on possible renormalizations see \cite{Yamada:2001px} and references therein):
\begin{equation} \label{eq:defshiftmix}
\delta \theta_{\tilde t}=\frac12 \frac{\hat\Pi_{12}(m_{\tilde t_1}^2)+\hat\Pi_{12}(m_{\tilde t_2}^2)}{m_{\tilde t_1}^2-m_{\tilde t_2}^2},
\end{equation}
where $\hat\Pi_{12}(p^2)$ is the off-diagonal self-energy of the stops. 
We define the OS combination $(m_t X_t)^{OS}$ from eqs.~(\ref{eq:defmix}) and (\ref{eq:defshiftmix}),
which implies
\begin{equation}
\frac{\delta (m_t X_t)}{m_t X_t}=\left(\frac{\delta m_{\tilde t_1}^2-\delta m_{\tilde t_2}^2}{m_{\tilde t_1}^2-m_{\tilde t_2}^2} +\frac{\delta \sin 2\theta_{\tilde t}}{\sin 2\theta_{\tilde t}} \right).
\label{eq:mtXtOS}
\end{equation}
In the usual definition for the stop mixing on-shell, $X_t^{\rm OS}=(m_tX_t)^{\rm OS}/m_t^{\rm OS}$, terms proportional to $\log(m_{\tilde t}/m_t)$ appear in the two loop thresholds. In the EFT approach these logs are big and need resummation. Therefore we use a different definition for $X_t$ which does not produce such terms and is more suitable for the EFT computation:
%We may be tempted to define $X_t^{OS}$ by $(m_t X_t)^{OS}/m_t^{OS}$. In this case if $X_t^{OS}$ is obtained directly from $X_t^{\overline{\rm DR}}$ the shift would introduce large logarithms $\log(m_{\tilde t}/m_t)$ in the two loop threshold corrections. If instead $X_t^{OS}$ is obtained from shifting $(m_t X_t)$ as in eq.~(\ref{eq:mtXtOS}) and then dividing by $m_t^{OS}$ the large logs
%are avoided but factors $m_t^{OS}/v^{\overline{\rm MS}}(m_{\tilde t})$ would appear.
%
%We opted for a third possibility. We define the stop mixing as
\begin{equation} \label{eq:defXOS}
X_t(Q)\equiv \frac{(m_t X_t)^{OS}}{m_t^{\overline{\rm MS}}(Q)}\,,
\end{equation}
where the numerator is computed from eq.~(\ref{eq:mtXtOS}) and we stress again that $m_t^{\overline {\rm MS}}(Q)$ is the top mass in the SM as any other $\overline{\rm MS}$ quantities in this paper.
%This definition has the advantages of avoiding the large logs and the extra 
%$m_t^{OS}/v^{\overline{\rm MS}}(m_{\tilde t})$ mentioned above.
An analogous definition applies for the sbottom and stau mixings. 
The decoupling of heavy particles like the gluino is ensured 
in our on-shell renormalization scheme.

For the squarks, the ${\cal O}(\alpha_s)$ shift (neglecting the quark masses)  reads \cite{Pierce:1996zz}:
\begin{equation}
\frac{\delta m_{\tilde q}^2}{m_{\tilde q}^2}=-\frac{g_3^2}{6\pi^2}\left[1+3x+(x-1)^2\ln|x-1|-x^2 \ln x+ 2 x \ln \frac{Q^2}{m_{\tilde q}^2} \right],
\end{equation}
with $x=M_3^2/m_{\tilde q}^2$. For the product $(m_t X_t)$ we obtain
\begin{align}
\delta(m_t X_t) &=\frac83 \frac{g_3^2}{(4\pi^2)} m_t \left[ 4 M_3- (2M_3-X_t)\log \frac{M_3^2}{Q^2}+ M_3\widetilde F_{10}\left(\frac{m_{Q_3}}{M_3} \right) 
+ M_3\widetilde F_{10}\left(\frac{m_{U_3}}{M_3}\right)   \right.
\label{eq:deltamtXtas}
\nonumber  \\
&+\left.  X_t \widetilde F_{11}\left(\frac{m_{Q_3}}{M_3},\frac{m_{U_3}}{M_3}\right)
\right]
\end{align}
and for the shift between the $\overline{\rm DR}$ top mass in the MSSM and the $\overline{\rm MS}$ top mass in the SM (which is also given in \cite{Bagnaschi:2014rsa})
\begin{equation}
 \frac{\delta m_t}{m_t} =-\frac{4}{3}\frac{g_3^2}{(4\pi)^2} \left[ 1+ \log\frac{M_3^2}{Q^2}+ \widetilde F_{6}\left(\frac{m_{Q_3}}{M_3} \right) 
+ \widetilde F_{6}\left(\frac{m_{U_3}}{M_3}\right)-\frac{X_t}{M_3} \widetilde F_{9}\left(\frac{m_{Q_3}}{M_3},\frac{m_{U_3}}{M_3}\right)
\right].
\end{equation}
%In our ``OS'' scheme the trilinear $A_t^{(\rm OS)}$ is scale dependent
%\begin{equation}
%\frac{d A_t^{(\rm OS)}}{dt}=\frac{1}{(4\pi)^2}\left[ \frac92 y_t^2-8 g_3^2 \right]X_t.
%\end{equation}
The functions $\widetilde F_{10}$ and $\widetilde F_{11}$ in eq.~(\ref{eq:deltamtXtas}) are defined as:
\begin{align}
 \widetilde F_{10}(x)&=\frac{1-x^2}{x^2}\ln|1-x^2|
\\ 
\widetilde F_{11}(x_1,x_2)&=-2+\frac{2(x_1^2\ln x_1^2-x_2^2\ln x_2^2)}{x_1^2-x_2^2}+\frac{x_1^2(1-x_2^2)^2\ln|1-x_2^2|-x_2^2(1-x_1^2)^2\ln|1-x_1^2|}{x_1^2 x_2^2 (x_1^2-x_2^2)}.
\end{align}

Analogously for the ${\cal O}(\alpha_t^2)$ corrections for degenerate scalars \cite{Espinosa:2000df}
\begin{align}
 \frac{\delta m_{\tilde t}^2}{m_{\tilde t}^2}=\frac{3 y_t^2}{32 \pi^2 s_\beta^2}&\biggl[(\hat{X}_t^2 s^2_\beta+\hat{Y}_t^2 c^2_\beta)\left(2-\ln \frac{m_{\tilde t}^2}{Q^2} \right) +c^2_\beta\left(1-\frac{\pi}{\sqrt{3}}\hat{Y}_t^2-\ln \frac{m_{\tilde t}^2}{Q^2}\right)
\nonumber \\
&+\hat{\mu}^4\ln{\hat{\mu}^2}+(1-\hat{\mu}^2)\left(3-2\ln \frac{m_{\tilde t}^2}{Q^2}\right)
-(1-\hat{\mu}^2)^2\ln(1-\hat{\mu}^2)
\biggr],
\\
\delta(m_t X_t)= \frac{3 y_t^2}{(4\pi)^2 s_\beta^2} &m_t\Biggl\{(X_t s^2_\beta+Y_t c^2_\beta)\left(2-\ln \frac{m_{\tilde t}^2}{Q^2}\right)
- \frac{\pi}{\sqrt 3}Y_t c_\beta^2+X_t\left(1-\frac32\ln \frac{m_{\tilde t}^2}{Q^2}\right) 
\nonumber \\
&\quad-\frac{1}{2} \bigl[1-\hat{\mu}^2+\hat{\mu}^4\ln{\hat{\mu}^2}+(1-\hat{\mu}^4)\ln(1-\hat{\mu}^2)  \bigr] X_t \Biggr\},
\end{align}
and the top mass shift is 
\begin{equation}
 \frac{\delta m_t}{m_t} =\frac{3}{4}\frac{y_t^2}{(4\pi)^2 s_\beta^2} \left[
 \left(1+c_\beta^2 \right)\left( \frac12-\ln \frac{m_{\tilde t}^2}{Q^2} \right)-\hat \mu^2 f_2(\hat \mu)\right].
\end{equation}
%In our ``OS'' definition $A_t$ runs to compensate the running of the SM $m_t$ in such a way that the combination $m_t X_t$ is scale independent.
%However, the combination $y_t A_t$ in the ``OS'' scheme is scale independent at this order in perturbation theory. 

As it was discussed in section~\ref{sec:on-shell}, in the $\overline{\rm DR}$ scheme there are power-like corrections from the gluino-stop loops to the Higgs quartic coupling which do not decouple in the limit of heavy gluino. We illustrate this effect for the simplified case of degenerate stops  
\begin{align}
\Delta \lambda^{(2, \, \overline{\rm DR})}_{\alpha_t \alpha_s}= \frac{y_t^4 g_3^2}{96\pi^4} & \left[12 \frac{M_3^2}{m_{\tilde t}^2}\left(1-\ln\frac{M_3^2}{Q^2}\right) -15+4\pi^2+12\ln\frac{M_3^2}{Q^2}-18\ln^2\frac{m_{\tilde t}^2}{Q^2}\right.
\nonumber \\
&\left.  -42\ln\frac{M_3^2}{m_{\tilde t}^2}  +12 \ln^2\frac{M_3^2}{m_{\tilde t}^2}+{\cal O}(M_3^{-1}) \right].
\end{align}
While our on-shell result, obtained from the $\overline{\rm DR}$ one by shifting the parameters in the one-loop ${\cal O}(\alpha_t)$ correction, guarantees the decoupling of heavy gluino
\begin{equation} \label{eq:heavygluinoOS}
\Delta \lambda^{(2,\, {\rm OS})}_{\alpha_t \alpha_s}= \frac{y_t^4 g_3^2}{96\pi^4}\left[
-30+4\pi^2+12\ln\frac{M_3^2}{Q^2}-18\ln^2\frac{m_{\tilde t}^2}{Q^2}-48 \ln\frac{M_3^2}{m_{\tilde t}^2}+12\ln^2\frac{M_3^2}{m_{\tilde t}^2}+{\cal O}(M_3^{-1})  \right].
\end{equation}
We also see that eq.~(\ref{eq:heavygluinoOS}) does not contain large logarithms $\ln m_{\tilde t}/m_t$. At last, the two-loop ${\cal O}(\alpha_t^2)$ SUSY threshold in our on-shell scheme is given by 
\begin{align}
\Delta \lambda^{(2)}_{\alpha_t^2}=&\frac{3 \,y_t^6 }{(4\pi)^4 s_\beta^2}\Bigg\{
\left(-10\ln \frac{m_{\tilde t}^2}{Q^2}+3\ln^2 \frac{m_{\tilde t}^2}{Q^2}\right)s_\beta^2 +
\frac{19}{2}-3{\hat{\mu}}^2-(8+2{\hat{\mu}}^2)f_1({\hat{\mu}})+3{\hat{\mu}}^2 f_2({\hat{\mu}})+4f_3({\hat{\mu}})
\nonumber \\ 
&+3\hat\mu^4\ln\hat\mu^4-(3-6\hat\mu^2+3\hat \mu^4)\ln(1-\hat\mu^2)+{\hat{X}}_t^6 \frac12 s_\beta^2 +{\hat{X}}_t^2\bigg[\frac{37}{2}+3\ln\frac{m_{\tilde t}^2}{Q^2}+9{\hat{\mu}}^2
\nonumber
\\
&+(4-6\hat \mu^2)f_1({\hat{\mu}})-4f_2({\hat{\mu}})-3{\hat{\mu}}^2 f_2({\hat{\mu}})
-9\hat \mu^4\ln\hat\mu^2-(3+6\hat\mu^2-9\hat\mu^4)\ln(1-\hat\mu^4)  \bigg]
\nonumber
\\
&-{\hat{X}}_t^4 \bigg[\frac{15}{4}+\frac32{\hat{\mu}}^2+\left(\frac12-\hat \mu^2\right)f_1({\hat{\mu}})
-\frac12 f_2({\hat{\mu}})-\frac32\hat\mu^4\ln\hat\mu^2-\left(\frac12+\hat\mu^2-\frac32\hat\mu^4 \right) \ln(1-\hat\mu^2)\bigg]
\nonumber
\\  
&+c^2_\beta\bigg[-\frac{7}{2}+60K+\pi^2-
{\hat{X}}_t^2\bigg(\frac{39}{2}+24K+3 \ln \frac{m_{\tilde t}^2}{Q^2}\bigg)+{\hat{X}}_t^4 \frac92
\nonumber
\\
&~~~~~~+{\hat{X}}_t{\hat{Y}}_t \left(12 -64K -4\sqrt3\pi\right)+{\hat{Y}}_t^2 \bigg(3-16K-\sqrt3 \pi \bigg)-\hat X_t^2\hat Y_t^2\left(\frac43-24K-\sqrt3 \pi  \right)
\nonumber
\\ 
&~~~~~~+{\hat{X}}_t^3 {\hat{Y}}_t \left(16K+\frac{2\pi}{\sqrt3}\right)
-{\hat{X}}_t^4{\hat{Y}}_t^2\bigg(\frac{7}{12} +8K + \frac{\pi}{2\sqrt3} \bigg)
\bigg]\Bigg\}.
\end{align}

\end{document}